\documentclass[prb,a4paper,twocolumn,floatfix]{revtex4-1}
\usepackage{color}
\usepackage[usenames,dvipsnames]{xcolor}
\usepackage{graphicx}
\usepackage{latexsym}
\usepackage[]{amsmath}
\usepackage{epstopdf}
\usepackage{amssymb}
\usepackage{array}

\begin{document}
\newcommand{\chem}[1]{\ensuremath{\mathrm{#1}}}

\title{Control of the third dimension in copper-based square-lattice antiferromagnets}

\author{Paul A. Goddard,$^{1*}$ John Singleton,$^{2,3}$ Isabel Franke,$^{3}$, 
Johannes S. M\"{o}ller,$^{3}$ Tom Lancaster,$^{4}$ Andrew J. Steele,$^{3}$ 
Craig V. Topping, $^{3}$ Stephen J. Blundell,$^{3}$ Francis L. Pratt,$^{5}$ C. Baines,$^{6}$
Jesper Bendix,$^{7}$ Ross D. McDonald.$^{2}$, Jamie Brambleby$^{1}$, Martin R. Lees$^{1}$, 
Saul H. Lapidus,$^{8}$  Peter W. Stephens,$^{8}$ Brendan W. Twamley,$^{9}$
Marianne M. Conner,$^{10}$  Kylee Funk,$^{11}$
Jordan F. Corbey,$^{10}$ Hope E. Tran,$^{10}$
J. A. Schlueter,$^{11}$ and Jamie L. Manson,$^{10*}$}

\affiliation{
$^{1}$ Department of Physics, University of Warwick, Gibbet Hill Road, Coventry, CV4 7AL, United Kingdom\\
$^{2}$National High Magnetic Field Laboratory, Los Alamos National Laboratory, MS-E536, Los Alamos, NM 87545, USA\\
$^{3}$Clarendon Laboratory, Department of Physics, University of Oxford, Parks Road, Oxford OX1~3PU, United Kingdom\\
$^{4}$Department of Physics, Durham University, South Road, Durham, DH1 3LE, United Kingdom\\
$^{5}$ ISIS Facility, STFC Rutherford Appleton Laboratory, Chilton, Didcot,
Oxfordshire, OX11~0QX, United Kingdom\\
$^{6}$Paul Scherrer Institut, Laboratory for Muon-Spin Spectroscopy, CH-5232 Villigen PSI, Switzerland\\
$^{7}$Department of Chemistry, University of Copenhagen, Copenhagen DK-2100, Denmark\\
$^{8}$Department of Physics and Astronomy, State University of New York, Stony Brook, NY 11794, USA\\
$^{9}$University Research Office, University of Idaho, Moscow, ID 83844, USA\\
$^{10}$Department of Chemistry and Biochemistry, Eastern Washington University, Cheney, WA 99004, USA\\
$^{11}$Materials Science Division, Argonne National Laboratory, Argonne, IL 60439, USA}

\begin{abstract}
Using a mixed-ligand synthetic scheme, we create a family of 
quasi-two-dimensional antiferromagnets, namely, 
[Cu(HF$_2$)(pyz)$_2$]ClO$_4$ [pyz = pyrazine], 
[Cu$L_2$(pyz)$_2$](ClO$_4$)$_2$ [$L$ = pyO = pyridine-N-oxide 
and 4-phpyO = 4-phenylpyridine-N-oxide. 
These materials are shown to possess equivalent 
two-dimensional [Cu(pyz)$_2$]$^{2+}$ nearly square layers, 
but exhibit interlayer spacings that vary from 6.5713~\AA ~to 
16.777~\AA, as dictated by the axial ligands. 
We present the structural and magnetic properties of this family 
as determined via x-ray diffraction, electron-spin resonance, 
pulsed- and quasistatic-field magnetometry and muon-spin rotation, 
and compare them to those of the prototypical two-dimensional 
magnetic polymer Cu(pyz)$_2$(ClO$_4$)$_2$. 
We find that, within the limits of the experimental error, the 
two-dimensional, {\it intralayer} exchange coupling 
in our family of materials remains largely unaffected by 
the axial ligand substitution, while the observed magnetic 
ordering temperature (1.91~K  for the material with the 
HF$_2$ axial ligand, 1.70~K for the pyO and 1.63~K for the 4-phpyO) 
decreases slowly with increasing layer separation.
Despite the  structural motifs common to this family and
Cu(pyz)$_2$(ClO$_4$)$_2$, the latter has significantly 
stronger two-dimensional exchange interactions and 
hence a higher ordering temperature. 
We discuss these results, as well as the mechanisms that might 
drive the long-range order in these materials, in terms of departures 
from the ideal $S=1/2$ two-dimensional 
square-lattice Heisenberg antiferromagnet.
In particular, we find that both spin exchange anisotropy
in the intralayer interaction and interlayer couplings
(exchange, dipolar, or both) are needed to account for the 
observed ordering temperatures, with the intralayer anisotropy
becoming more important as the layers are pulled further apart.
\end{abstract}
\maketitle
\section{Introduction}

As a quintessentially quantum-mechanical model and because of its
relevance to the unsolved problem of high-temperature superconductivity, 
the $S=1/2$ two-dimensional (2D) square-lattice Heisenberg
antiferromagnet (SLHAFM) remains one of the most actively studied 
systems in condensed matter physics~\cite{lines,chakravarty88,chakravarty89,tyc89,manousakis}. 
It is described by the Hamiltonian 
$J \Sigma_{ij}{S_i \cdot S_j}$ (where $J$ is the strength of the nearest-neighbor 
exchange interaction) and, because of strong thermal and 
quantum fluctuations, is resistant to long-range order for temperatures
$T > 0$~K~[\onlinecite{mermin}]. 
Considerable theoretical attention has been paid to additions to this 
ideal model, including external magnetic fields, next-nearest-neighbor 
coupling, or departures from isotropy in the interactions. 
Predicted consequences of such deviations include alterations 
to the excitation spectra~\cite{syljuasen02}, 
changes in the universality class~\cite{cuccoliprb03}, 
the emergence of exotic magnetic phases~\cite{chandra88}, 
and the promotion of a variety of ordered states~\cite{sushkov01}. 

Practical realizations of the SLHAFM involve
crystalline materials that are inevitably three dimensional
from a structural standpoint~\cite{njp};
this introduces possible interlayer magnetic coupling
between adjacent square-lattice planes, 
which, if present, will raise the magnetic ordering temperature 
to non-zero values and, potentially, 
obscure the effects of the other perturbations~\cite{njp}. 
Reducing the interlayer coupling may allow the above effects to 
come to the fore, permitting the associated theoretical 
predictions to be investigated in the laboratory. 
In this context, the closest experimental approximations to ideal 2D magnets
have  typically made use of copper-oxygen superexchange 
bonds~\cite{lancasterprb07}. 
However, crystal-engineering efforts have given 
rise to a number of reduced-dimensionality molecular candidates~\cite{mansonag,steele,selmani10} in which the 
interlayer coupling is reduced to levels seen in the best 
inorganic materials. 
In this way it has been possible to show 
evidence of changes in universality class as a function of magnetic 
field in a molecular SLHAFM~\cite{kohama10,fortune2014}. 
An added advantage of molecular systems is the ability to make 
controlled adjustments to the crystal structure, 
thereby tuning interaction strengths and better testing the 
predictions of the SLHAFM and associated models. 
To this end, we and others have previously shown that it is 
possible to gain a degree of control over the primary exchange 
energy in low-dimensional molecular antiferromagnets via constitutional 
changes that include deuteration~\cite{isotope}, 
anion substitution~\cite{njp,woodward1}, exchange of 
halide ligands~\cite{butcher09,lapidus13}, and the application of high pressures~\cite{halder,saman13}. 

In this paper, 
we examine a family of materials based on 2D arrays 
of magnetic Cu(II) ions linked by pyrazine (C$_4$H$_4$N$_2$) 
ligands in order to  investigate the effects of tuning the interlayer 
coupling strength. 
By varying the layer separation 
through the interchange of axial ligands, 
we study to what extent the antiferromagnetic 
ordering temperature can be manipulated, and whether it is possible to  
approach the ideal 2D limit. 
The four coordination polymers chosen are: 

({\bf 1})~[Cu(HF$_2$)(pyz)$_2$]ClO$_4$, 

({\bf 2})~[Cu(pyO)$_2$(pyz)$_2$](ClO$_4$)$_2$, 

({\bf 3})~[Cu(4-phpyO)$_2$(pyz)$_2$](ClO$_4$)$_2$, and 

({\bf 4})~Cu(pyz)$_2$(ClO$_4$)$_2$.\\
Here pyz = pyrazine, pyO = pyridine-$N$-oxide (C$_5$H$_5$NO), 
and 4-phpyO = 4-phenyl-pyridine-$N$-oxide (C$_{11}$H$_9$NO). 
These materials have interlayer spacings that range from 6.5713 to 16.777~\AA~at low temperatures. {\bf 4} is the prototypical example of a molecular SLHAFM~\cite{lancasterprb07,woodward1,darriet} and is used here as a 
yardstick by which to judge the other compounds.
 
This paper is organized as follows. 
After outlining the synthesis procedures, experimental details
and calculations 
in the next two sections, we describe the crystal structure of the materials 
as determined using x-ray diffraction. 
Next we present the results of electron spin-resonance 
and pulsed-field magnetization experiments to determine the 
anisotropic $g$-factors and effective nearest-neighbor 
exchange energies, respectively. 
We describe how the exchange energies compare with those found 
from previous measurements on structurally related materials and 
examine possible reasons for their variation across the family 
using density-functional theory. 
Then we present the determination of the magnetic 
ordering temperatures from muon-spin rotation measurements, 
followed by a discussion of the nature of the magnetic 
interactions between the layers, as well as the mechanism 
that drives these systems to long-range order at low temperatures.

\section{Experimental details}
\subsection{Sample synthesis} 
All chemical reagents were obtained from commercial sources 
and used as received. 
Cu(ClO$_4$)$_2\cdot$6H$_2$O served as the Cu(II) source in all syntheses. 
For {\bf 1}, Cu(II) ions were slowly mixed with an aqueous 
solution containing stoichiometric amounts of 
NH$_4$HF$_2$ (0.0627~g, 1.10~mmol) and 
pyrazine (0.1762~g, 2.20~mmol) to afford a blue solution. 
For {\bf 2} and {\bf 3}, an aqueous solution of Cu(II) was combined 
with an ethanolic solution containing a mixture of 
pyrazine (0.500~g, 2.11~mmol) and pyridine-N-oxide (0.401~g, 4.20~mmol) 
or 4-phenylpyridine-N-oxide (0.7190~g, 4.20~mmol). 
Compound {\bf 4} was synthesized as described in the 
literature~\cite{woodward1}. 
Deep-blue ({\bf 1} and {\bf 4}) or blue-green ({\bf 2} and {\bf 3}) 
solutions were obtained and, when allowed to slowly 
evaporate at room temperature, x-ray quality blue plates 
({\bf 1} and {\bf 4}) or dark green plates ({\bf 2}) 
$\sim 0.5 \times 0.5 \times 0.2$~mm$^3$ were recovered, whereas {\bf 3} 
yielded dark green microcrystals $\sim0.03 \times 0.03 \times 0.01$~mm$^3$. 
The relative amounts of pyz:pyO ({\bf 2}) or 
pyz:4-phpyO ({\bf 3}) were optimized in order to 
prevent formation of byproducts such as 
Cu(ClO$_4$)$_2$(pyz)$_2$, [Cu(pyO)$_6$](ClO$_4$)$_2$, 
or [Cu(4-phpyO)$_6$](ClO$_4$)$_2$. 

\subsection{X-ray diffraction studies}
Crystals of each compound except 
[Cu(pyz)$_2$(4-phpy-O)$_2$](ClO$_4$)$_2$ were carefully selected, 
attached to a glass fiber and data collected at several 
temperatures between 100 and 297~K using a 
Bruker APEX II CCD X-ray diffractometer (Mo-K$\alpha$ radiation,  
$\lambda= 0.071073$~nm) equipped with a low-temperature device. 
Measurements employed omega scans and a 
full sphere of data was collected. 
Cell parameters were retrieved using SMART software~\cite{smart}, 
and data were refined using SAINTPlus~\cite{saintplus} 
based on all observed reflections. 
Data reduction and correction for Lorentz polarization and 
decay were performed using the SAINTPlus software. 
Absorption corrections were applied using SADABS~\cite{sadabs1}.  
Structures were solved directly and refined by the least-squares 
method on F$^2$ using the SHELXTL program package~\cite{shelxtl}.  
All non-hydrogen atoms were refined anisotropically. 
No decomposition was observed during data collection. 

High-resolution synchrotron X-ray powder diffraction patterns 
were collected on [Cu(pyz)$_2$(4-phpy-O)$_2$](ClO$_4$)$_2$
at 297~K using the 11-BM-B beamline located at the 
Advanced Photon Source, Argonne National Laboratory. 
X-rays were selected using a Si(111) channel cut monochromator. 
After the sample, the diffracted beams were analyzed using 
a Ge(111) crystal and detected by a NaI scintillation counter. 
The wavelength and diffractometer zero were calibrated using a sample of 
NIST Standard Reference Material 1976, a sintered plate of Al$_2$O$_3$. 
The sample was loaded into a 1.0~mm diameter Kapton tube 
and mounted in a sample automation robot. 
Data were collected for approximately 1~hour. 
To improve particle statistics, the capillary was spun at several 
radians per second.
Results of the data refinement for all four
materials are given in Table~\ref{lattice} below. 

\subsection{Electron-spin resonance}
Electron-spin resonance spectra were measured on single-crystal samples 
of {\bf 1} and {\bf 2}, and fine powders of {\bf 3} and {\bf 4} in 
the frequency range 10 to 110~GHz using cavity perturbation techniques. 
For single-crystal $g$-factor anisotropy measurements, 
a monomoded cavity, resonating at a frequency of around 71~GHz 
and mounted on a cryogenic goniometer was employed~\cite{schramaJPCM}.  
This allows the crystal to be rotated with respect to the applied 
magnetic field without thermal cycling. 
For powder samples, and to examine the frequency-field scaling 
for single crystals, over-moded cylindrical 
and confocal resonators were used. 
Temperature control was provided by a standard 
$^4$He flow cryostat and/or a single shot $^3$He refrigerator. 
Quasistatic magnetic fields were applied using a 17~T superconductive solenoid; 
a Millimetre-wave Vector Network Analyzer (MVNA), 
manufactured by AB-Millimetre, was used as 
both the microwave source and detector~\cite{schramaJPCM}. 

\subsection{Magnetometry} 
Pulsed-field magnetization measurements
were performed  at the National High Magnetic Field 
Laboratory in Los Alamos; fields of up to
65~T with typical rise times $\approx 10$~ms were used. 
Single crystals are mounted in 
1.3~mm diameter PCTFE ampoules (inner diameter 1.0~mm) 
that can be moved into and out of a 1500-turn, 1.5~mm bore, 
1.5~mm long compensated-coil susceptometer, 
constructed from 50~gauge high-purity copper wire~\cite{njp}. 
When the sample is within the coil and the field pulsed 
the voltage induced in the coil is proportional to the rate of 
change of magnetization with time, $({\rm d} M/{\rm d}t)$. 
Accurate values of the magnetization are obtained by numerical
integration of the signal with respect to time, followed by
subtraction of the integrated signal recorded using an empty 
coil under the same conditions~\cite{njp}. 
The magnetic field is measured via the signal induced within a coaxial 
10-turn coil and calibrated via observation of de Haas--van Alphen 
oscillations arising from the copper coils of the susceptometer~\cite{njp}. 
The susceptometer is placed inside a $^3$He cryostat, 
which can attain temperatures as low as 500~mK.
During each experiment, the size and sign of ${\rm d}H/{\rm d}t$ 
of the field pulses was varied. 
No evidence of a hysteresis caused by slow relaxation 
of the sample moment was observed in any of measurements.
 
Low-field magnetization at temperatures $\approx 500$~mK 
was measured using 
an iQuantum low-temperature insert for the Quantum Design 
Magnetic Property Measurement System (MPMS) XL SQUID magnetometer. 
Zero-field cooled magnetic-moment measurements in the field range 
$0\le \mu_{0}H \le 0.3$~T were performed on 
76.007, 47.326 and 28.717~mg polycrystalline samples of 
\textbf{1}, \textbf{2} and \textbf{3}, respectively. 
To make the measurements, the sample is placed inside a 
polycarbonate capsule with a small amount of cotton wool 
to prevent it from moving. 
The capsule is wrapped in a thermally conducting sheath,
containing copper wires arranged parallel to the magnetic
field, and 
fixed inside a plastic drinking straw. 
The temperature is monitored via a thermometer positioned
approximately 1~cm  above the sample inside the straw. 
The straw is then mounted to the end of a rod which is lowered into the
$^{3}$He cryostat. 
With the $^{4}$He chamber of the MPMS cooled to 1.6~K, 
the cryostat is evacuated and liquid $^{3}$He is allowed to condense inside,
such that the sample is submerged. 
Temperatures down to 500~mK are achieved by combined use of a 
turbo~molecular pump and a charcoal sorption pump. 

\subsection{Muon-spin rotation}
Zero-field muon-spin rotation measurements were performed on 
powder samples of  {\bf 2} and {\bf 3}. 
Sample {\bf 2} was covered by a 25~$\mu$m silver foil 
and mounted on the cold finger of the dilution refrigerator on 
the LTF instrument at the Swiss Muon Source, 
Paul Scherrer Institut, Switzerland. 
Sample {\bf 3} was covered by a 12.5~$\mu$m silver foil and 
mounted in a $^3$He cryostat on the ARGUS instrument at the ISIS facility, 
Rutherford Appleton Laboratory, UK. 
Further details of muon-spin rotation experiments on 
{\bf 1} and {\bf 4} can be found in 
References~\onlinecite{lancasterprb07} and~\onlinecite{steele}, respectively.

\section{Calculations}
\subsection{Density functional theory}
The broken-symmetry approach of Noodleman~\cite{noodleman} 
as implemented in the ORCA version 2.8 suite of 
programs~\cite{neese1,neese2,sinnecker} was employed
to evaluate the exchange couplings. 
The quoted coupling constants are based on formalism of Yamaguchi, 
which employs calculated expectation values $\langle S^2\rangle$ 
for both high-spin and broken-symmetry states~\cite{yamaguchi,soda}. 
Calculations employed the PBE0 functional, which has previously been 
demonstrated to yield reliable values for magnetic couplings in analogous systems~\cite{lapidus13,manson11}. 
The Ahlrichs-VTZ basis function set was used~\cite{ahlrichs}. 
For compound {\bf 3}, SCF convergence was facilitated by using a 
non-standard value (10) for the DIISMaxEq parameter in ORCA. 

\subsection{Dipole-field calculations}
The dipolar interaction of a spin with magnetic moment 
${\bf m}_0$ at position ${\bf r}_{0}$ with the 
magnetic moment ${\bf m}_i$ (assumed to be completely localized) 
of ion $i$ at position ${\bf r}_i$ is given by  
\begin{equation}
B_{\rm dip}^{\alpha}({\bf r}_{0})=\sum_i D^{jk}_i ({\bf r}_{0}) m_i^k,
\end{equation}
where 
\begin{equation}
D_i^{jk}=\frac{\mu_0}{4\pi R_i^3}(\frac{3R_i^j R_i^k}{R_i^2}-\delta^{jk})
\end{equation}
is the dipolar tensor with indices $jk$
and ${\bf R}_i=(R_i^x,R_i^y,R_i^z)={\bf r}_0-{\bf r}_i$. 

The dipolar tensor is evaluated for the bulk crystal inside a 
Lorentz sphere of radius $r_{\rm L}=100$~\AA. 
The results are well converged due to the relatively short-range 
nature of the dipolar interaction. 
The interlayer dipolar interaction was estimated by calculating 
the intralayer dipolar interaction inside a Lorentz circle of 
radius $r_{\rm L}=100$~\AA\ and subtracting the result 
from the bulk 3D dipolar interaction. 
The dipolar energy of a chosen moment ${\bf m}_0$ is then given by 
\begin{equation}
E_{\rm dip}=-{\bf m}_0 \cdot {\bf B_{\rm dip}}
\label{eqn:dipen}
\end{equation}
and scales with the square of the ordered moment size ${\bf m}_i^2$. 

This calculation was performed by assuming certain collinear 
magnetic structures (see below) and the energy in Eq.~\ref{eqn:dipen} 
was then calculated for many different directions of the magnetic moments. 
We define $E_{\rm D}=(E^{\rm max}_{\rm dip}-E^{\rm min}_{\rm dip})/2$, 
where $E^{\rm min}_{\rm dip}$ ($E^{\rm max}_{\rm dip}$) is the 
minimum (maximum) dipolar energy given the considered magnetic structure.
\begin{figure*}[ht]
\centering
\includegraphics[width=18.0cm]{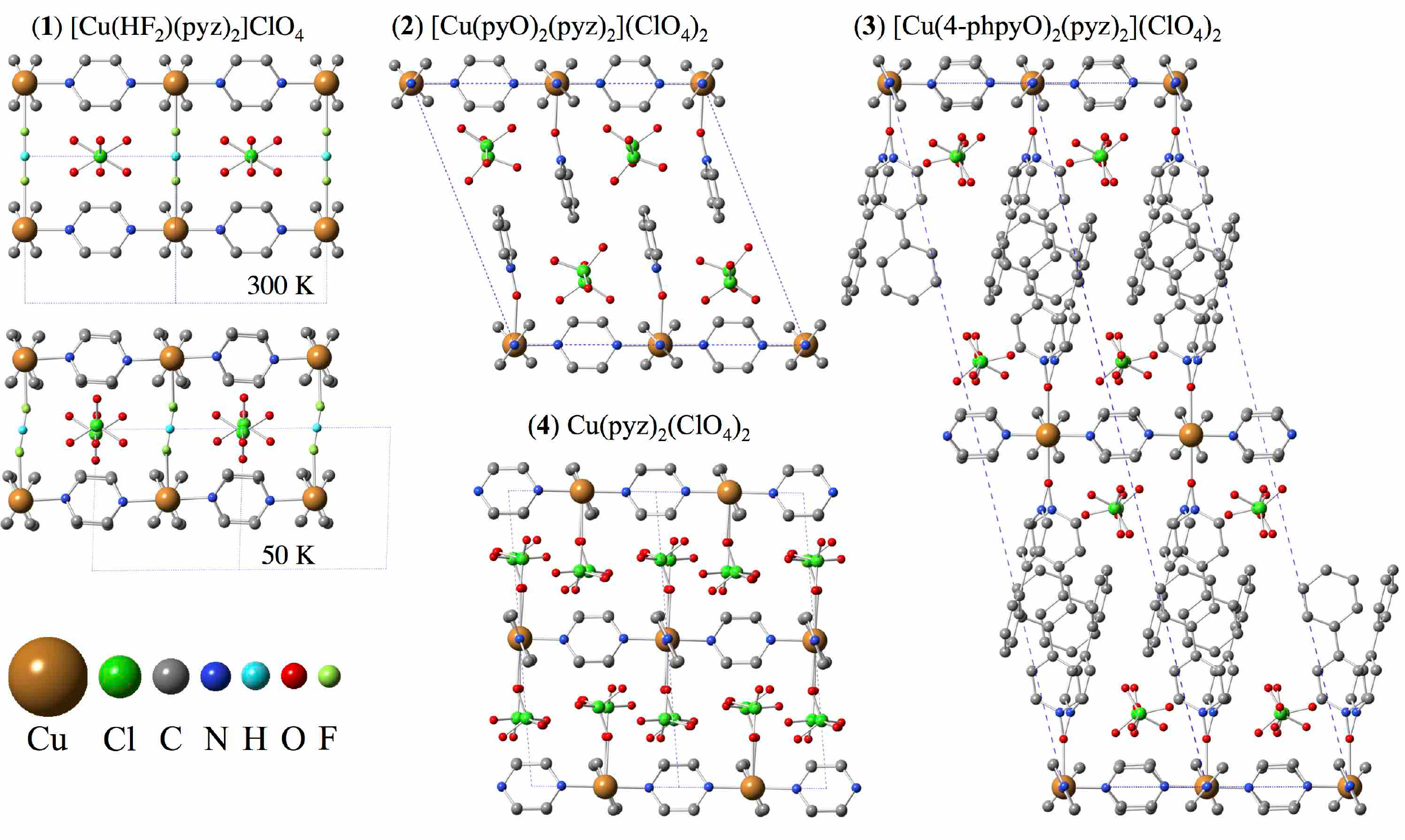}
\caption{(color online). Experimentally determined crystal structures viewed along the copper-pyrazine planes.
All structures shown were determined via single-crystal 
x-ray diffraction at 100~K, except for {\bf 1} for which 
both 300~K and 50~K structures are presented. 
Here pyz = pyrazine, pyO = pyridine-N-oxide, 
4-phpyO = 4-phenyl-pyridine-N-oxide;
as shown in the inset key,
Cu = brown, Cl = green, C = gray, N = blue, H = cyan, O = red, 
and F = light green. 
All hydrogens from organic molecules are omitted for clarity. 
The blue-dashed lines denote a single unit cell. 
All four systems are based on two-dimensional arrays of 
Cu-pyz in which pyz orbitals 
mediate the dominant exchange interactions. 
In {\bf 1} the triclinic structure is supported by strong 
F$\cdots$H$\cdots$F hydrogen bonds that form bridging 
ligands separating the layers~\cite{njp}. 
The three other systems have non-bridging 
ligands along the interlayer direction and a monoclinic structure. 
The non-bridging ligands in {\bf 2} and {\bf 3} are pyO  
and 4-phpyO respectively. Whilst the ClO$_4$ molecules are 
non-coordinating anions in {\bf 1}, {\bf 2} and {\bf 3}, 
they coordinate to the Cu-ions in {\bf 4}. 
}
\label{strucfig}
\end{figure*}

\renewcommand{\arraystretch}{1.3}
\begin{table*}[t]
\centering
\newcolumntype{C}{ >{\centering\arraybackslash} m{1.3cm} }
\newcolumntype{D}{ >{\centering\arraybackslash} m{1.6cm} }
\begin{tabular}{m{4.95cm}*{2}{C}*{3}{D}*{1}{C}*{1}{D}*{1}{C}}
\hline \hline
& $T$ (K) & space group & $a$ (\AA) & $b$ (\AA) & $c$ (\AA) & $\alpha$ ($^\circ$) & $\beta$ ($^\circ$) & $\gamma$ ($^\circ$) \\
\hline
({\bf 1}) [Cu(HF$_2$)(pyz)$_2$]ClO$_4$ & 300 & $P4/nmm$ &  9.7054(6) & 9.7054(6) & 6.6894(9) & 90.00 & 90.00 & 90.00 \\
 & 50 & $P\bar{1}$ &   6.5943(1) & 9.6300(1) & 9.7089(2) & 90.0006(4) & 94.791(1) & 91.720(1)\\
\hline
({\bf 2}) [Cu(pyO)$_2$(pyz)$_2$](ClO$_4$)$_2$ & 300 & $C2/m$ &  13.7154(2) &13.7014(2) & 13.1926(2) & 90.00 & 108.637(1) & 90.00\\
 & 100 & $C2/m$ &  13.6676(2) &13.6699(2) & 13.1910(2) & 90.00 & 111.572(1) & 90.00 \\
\hline
({\bf 3}) [Cu(4-phpy-O)$_2$(pyz)$_2$](ClO$_4$)$_2$ & 300 & $C2/c$ & 34.566(2) & 9.4850(3) & 9.8863(3) & 90.00 &  102.763(3) & 90.00\\
 & 100 & $C2/c$ &35.6195(4) & 9.44905(9) & 9.83656(7) & 90.00 &  109.606(1) & 90.00\\
\hline								 																	
({\bf 4}) Cu(pyz)$_2$(ClO$_4$)$_2$ & 300 & $C2/m$ & 9.734(2) & 9.729(2) & 8.132(2)  & 90.00 & 120.855(4) & 90.00\\
 & 100 & $C2/c$ & 13.9276(3) & 9.7438(2) & 9.7871(2)  & 90.00 & 96.924(1) & 90.00\\
\hline \hline
\end{tabular}
\caption{Space groups and lattice parameters of the compounds in Figure~\ref{strucfig}. The 300~K data for {\bf 4} are taken from \cite{woodward1}; the rest are from this study.}
\label{lattice}
\end{table*}

\renewcommand{\arraystretch}{1.3}
\begin{table*}[t]
\centering
\newcolumntype{C}{ >{\centering\arraybackslash} m{1.5cm} }
\newcolumntype{D}{ >{\centering\arraybackslash} m{1.3cm} }
\begin{tabular}{m{5cm}*{1}{C}*{2}{C}*{4}{C}}
\hline \hline
										
& equatorial Cu--N \linebreak (\AA) 		& axial Cu--$L$ \linebreak (\AA)   	& intralayer Cu-Cu (\AA) 				
& interlayer Cu-Cu (\AA) 	& interlayer separation (\AA)  & pyrazine tilt ($^\circ$)			& pyrazine twist \linebreak ($^\circ$)\\
\hline
({\bf 1}) [Cu(HF$_2$)(pyz)$_2$]ClO$_4$ & 2.064(5) 2.075(5) 2.078(5) 2.089(6)	& 2.312(4)			& 6.824(1) 6.829(1) 6.846(1) 6.850(1) 	& 6.5943(1) & 6.5713(1)				
& 52.8 \linebreak 55.9 \linebreak 67.1 \linebreak 67.7 	
& 3.11 \linebreak 3.58 \linebreak 5.37 \linebreak 7.41\\
\hline
({\bf 2}) [Cu(pyO)$_2$(pyz)$_2$](ClO$_4$)$_2$ 	& 2.031(1) 2.0315(9) 2.0316(9)			& 2.3148(8)	 				& 6.830(1) 6.834(1) 6.840(1)			& 12.426(2) 13.191(2)	& 12.267(2)				& 47.3 \linebreak 53.4 \linebreak 68.1 				& 0.00 \linebreak 0.48\\
\hline
({\bf 3}) [Cu(4-phpy-O)$_2$(pyz)$_2$](ClO$_4$)$_2$ & 2.009(5) 2.019(5)					& 2.292(2)						& 6.812(1) 6.827(1)					& 16.81(1)				& 16.777(9)				& 50.6 \linebreak 64.4 							& 1.40 \linebreak 2.31\\
\hline								 																	
({\bf 4}) Cu(pyz)$_2$(ClO$_4$)$_2$ 			& 2.057(1) 2.058(1)					& 2.356(9)						& 6.904(1) 6.907(1)					& 8.014(1)				& 6.913(1)					& 63.0 \linebreak 69.9 							& 1.08 \linebreak 1.97\\
\hline								 																	
~~~~ [Cu(HF$_2$)(pyz)$_2$]SbF$_6$ 				& 2.033(1)							& 2.259(1)						& 6.836(1) 						& 6.785(1)				& 6.785(1)					& 79.2 		 								& 0.00\\
\hline \hline
\end{tabular}
\caption{Crystallographic parameters of interest for 
magnetism, measured at 100~K, except for {\bf 1} for which the 
data were taken at 50~K. 
Cu--N represents the equatorial coordination bond length and 
Cu--$L$ is the axial coordination bond length, {\it i.e.} Cu--F for {\bf 1}
and Cu--O otherwise. 
In all cases the Jahn-Teller axis lies along the axial ligand direction. 
Intralayer Cu-Cu are the Cu--pyz--Cu distances within the planes. 
Interlayer Cu-Cu are the shortest interlayer neighbor distances; 
where more than one value is quoted it is because they are closely 
spaced due to staggering of adjacent planes. 
The interlayer spacings are the perpendicular distances 
between planes. 
The tilt angle is that between the plane of the pyrazine 
molecules and the copper-pyrazine layers. 
The twist angle is that between the N$\cdots$N 
axis of the pyrazine molecules and the copper-copper 
pathways in the two-dimensional planes. 
The last line shows data measured at 15~K 
for the related material [Cu(HF$_2$)(pyz)$_2$]SbF$_6$~[\onlinecite{njp}],
which is compared to {\bf 1-4} in later discussions.
}
\label{structab}
\end{table*}

\section{Crystal structures}
The low-temperature crystal structures of all four compounds 
viewed along the copper-pyrazine planes are shown 
in Figure~\ref{strucfig}; complete data sets are available
at the Cambridge Crystallographic Data Centre (CCDC)~\cite{CCDC}.
Table~\ref{lattice} collects the lattice parameters at both high 
and low temperatures, while Table~\ref{structab} offers a 
comparison of some low-temperature structural parameters 
relevant to the discussion of the magnetic properties. 

[Cu(HF$_2$)(pyz)$_2$]ClO$_4$ ({\bf 1}) crystallizes at room temperature 
in the tetragonal space group $P4/nmm$, with each copper(II) ion 
equatorially ligated to four pyrazine nitrogen atoms at a 
distance of 2.039(3)~\AA ~and axially ligated to the fluorines 
from HF$_2^-$ at a distance of 2.228(3)~\AA. 
This coordination gives rise to square-lattice 
copper-pyrazine layers separated by strong 
hydrogen-bonded HF$_2^-$ pillars. 
Intralayer and interlayer copper separations are 
6.863(1) \AA~and 6.689(1)~\AA, respectively, with the 
non-coordinated ClO$_4^-$ counterions located close to the 
centre of these copper-cornered cuboids.  
The pyrazine molecule is tilted 64.2$^\circ$ out of 
the copper-pyrazine planes and counter-rotates relative 
to its closest neighbor within the plane. 
This room-temperature structure has been 
reported previously~\cite{njp}. 
On cooling to 50~K the structure transforms into the triclinic 
space group $P\bar{1}$. 
The Cu--N bond lengths are no longer equal, varying 
between 2.064(5) and 2.089~\AA. 
This leads to a slight departure from perfect copper-pyrazine 
square planes, the shortest in-plane Cu$\cdots$Cu distance 
being 6.824(1)~\AA~and the longest 
6.850(1)~\AA~(see Table~\ref{structab}). 
The pyrazine molecules are still counter-rotated relative 
to their neighbors, but now there are four different tilt angles, 
ranging from 52.8$^\circ$ to 67.7$^\circ$. 
In addition, while the N$\cdots$N axis of each pyrazine 
molecule in the room-temperature structure lies along the path 
which connects the copper ions within the planes, at low 
temperatures the molecules undergo a twist such that 
the N$\cdots$N axes make an angle with this pathway 
that varies between 3.11$^\circ$ to 7.41$^\circ$. 

This material is part of the closely related family 
[Cu(HF$_2$)(pyz)$_2$]$X$, where the local symmetry of counterion 
$X$ can be octahedral ({\it e.g.} PF$_6^-$, AsF$_6^-$, 
SbF$_6^-$, NbF$_6^-$ and TaF$_6^-$) or tetrahedral 
({\it e.g.} ClO$_4^-$, BF$_4^-$)~\cite{njp}. 
The main variation in the room temperature structure on 
moving across this family is in the tilt angle of the pyrazines. 
For the materials with octahedral anions, this angle 
is close to 90$^\circ$, but is significantly smaller for 
the tetrahedral-anion materials~\cite{njp}. 
More differences appear on cooling. 
In contrast to the 50~K structure of the $X =$ ClO$_4^-$ 
material described here, the structure of the $X =$ SbF$_6^-$ material 
measured at 15~K indicates that this system remains 
in a tetragonal space group down to low-temperatures, 
has a single pyrazine tilt angle of 79.2$^\circ$, 
and does not exhibit the twisting of the 
N$\cdots$N pyrazine axis mentioned above. 
In addition, as discussed in more detail below, 
there is a reduction of nearly a factor of two in the 
primary exchange constant in this family associated 
with the change from octahedral to tetrahedral anions~\cite{njp}.

[Cu(pyO)$_2$(pyz)$_2$](ClO$_4$)$_2$ ({\bf 2}) adopts the 
monoclinic $C2/m$ space group with no change on cooling to 100~K. 
There is a similar equatorial ligand arrangement in the
low-temperature phase of {\bf 1}, 
with Cu--N distances ranging from 2.023(3) to 2.040(4)~\AA~at 
low temperatures and copper-pyrazine plaquettes that deviate 
slightly from a square arrangement. 
Each copper(II) site is also coordinated to two oxygens from 
the axially-ligated pyridine-N-oxide molecules at a distance 
of 2.331(3)~\AA ~and the bond angles within the 
CuN$_4$O$_2$ octahedra that deviate from 
90$^\circ$ by up to 2.6$^\circ$. 
Adjacent copper-pyrazine planes are shifted with 
respect to one another along the $a$-axis, 
leaving each copper close to equidistant from its two 
nearest neighbors within each adjacent plane. 
This staggering of layers could give rise to a degree of frustration 
if any antiferromagnetic interlayer exchange coupling is present, 
and in principle cause complete cancellation if the 
copper ions were arranged in a
triangular lattice in the interlayer 
direction. In reality there is a difference in the interlayer 
Cu--Cu distances, as shown in Table~\ref{structab}. 
As compared to  {\bf 1}, {\bf 2} has twice as many ClO$_4^-$ anions 
per formula unit, with the chlorine ion approximately centrally
located above and below the copper-pyrazine plaquettes at alternating 
distances of 3.0 and 3.5~\AA.
 
[Cu(4-phpyO)$_2$(pyz)$_2$](ClO$_4$)$_2$ ({\bf 3}) also has a 
monoclinic structure, crystallizing in the $C2/m$ space group. 
The axial ligand in this material is the larger 
4-phenyl-pyridine-N-oxide molecule. 
The ligand and anion arrangements are akin to that of {\bf 2} with 
departures from the ideal square lattice in the copper-pyrazine 
planes and a similar range of bond angles within 
the CuN$_4$O$_2$ octahedra. 
Again there are two non-coordinated counterions 
per formula unit, the chlorines here lying at a distance of 
approximately 3.5~\AA ~either side of the copper-pyrazine planes.

The structure of Cu(pyz)$_2$(ClO$_4$)$_2$ ({\bf 4}) has been 
reported by Darriet {\it et al.}~\cite{darriet} and more recently by
Woodward {\it et al.}~\cite{woodward1}. 
The structure resulting from our x-ray diffraction study is 
shown in Figure~\ref{strucfig} and is in agreement with the 
previous measurements. 
At room temperature the crystal adopts the monoclinic 
$C2/m$ space group with doubly-staggered square-lattice 
copper-pyrazine planes and a single pyrazine tilt angle of 65.8$^\circ$. 
Below about 180~K a phase change to the $C2/c$ space group occurs. 
The square layers remain, but there are now two tilt angles, 
63.0$^\circ$ and 69.9$^\circ$, one for each 
$cis$-coordinated ligand pair~\cite{woodward1}. 
The tilt angles do not counter-rotate within the copper-pyrazine planes. 
Unlike compounds {\bf 1}, {\bf 2} and {\bf 3}, the ClO$_4^-$ ions are 
coordinated to the coppers via an oxygen. 
At temperatures below the phase change, these ligands order 
such that the chlorines are a perpendicular distance of 
3.7~\AA ~away from the planes to which their molecules 
coordinate and sit in the space above the copper-pyrazine 
squares of the adjacent plane at a distance of 3.2~\AA.

Thus, by changes in the axial coordination of these four materials,
a family is realized in which the copper-pyrazine planes are largely 
maintained while the interlayer structure is significantly altered. 
In particular, the spacing between adjacent layers at low 
temperatures varies from 6.5713(1)~\AA, through 12.267(2)~\AA, 
to 16.777(9)~\AA~for the systems with the 
HF$_2$ ({\bf 1}), pyO ({\bf 2}) and 4-phpyO ({\bf 3}) ligands, respectively. 
System {\bf 4}, with the ClO$_4$ axial ligands, has an interlayer spacing of 6.913(1)~\AA.

\section{Single-ion properties}
In all four materials the Jahn-Teller-active copper(II) ion sits at the center 
of a distorted CuN$_4$$L_2$ octahedron, where $L$ is the 
axially-coordinated ion, namely F in compound {\bf 1}, 
and O in the other compounds. 
As shown in Table~\ref{structab}, in each case the Cu--$L$ 
bond length is elongated compared to the intralayer Cu--N 
coordination bond lengths, an indication that the Jahn-Teller 
axis points out of the copper-pyrazine planes~\cite{EinM}. 
For $S=1/2$ copper in a such an environment, 
the crystal field splits the $3d$ states such that the
d$_{x^2-y^2}$ orbital is partially occupied and aligned 
perpendicular to the axial distortion~\cite{EinM}. 
The greater overlap of  d$_{x^2-y^2}$ with the d$_{xy}$ orbital, 
than with d$_{xz}$ or d$_{yz}$, 
leads to a greater enhancement of the $g$-factor for fields along 
$z$ (the axial direction) than for those in the $xy$ plane~\cite{Lancaster2014}. 
Similarly, through the spin-orbit interaction, 
the exchange coupling between neighboring spins will likely 
also possesses spin anisotropy~\cite{EinM}. 
This spin exchange anisotropy (as opposed to the spatial exchange 
anisotropy dictating the dimensionality of the magnetism) is 
a two-ion anisotropy, which leads to easy-plane or easy-axis 
deviations from Heisenberg (spin isotropic) antiferromagnetic order.
As discussed later, the family of quasi-two-dimensional 
$3d^9$ copper systems considered here exhibits small levels of 
spin anisotropy, and which are likely, at least in part, 
responsible for the long-range order observed in these materials.

For a spin-half system in the absence of interactions 
between the spins, the electron-spin resonance (ESR) 
condition (the frequency-field relationship for the ESR line) 
is a measure of the spectroscopic g-factor~\cite{Abragam,turovbook}. 
The ESR condition in the antiferromagnetic state is 
determined by both the applied magnetic field and 
the effective internal exchange field~\cite{Knaflic,Dolinsek}. 
This has the largest effect on the frequency-field scaling, 
and hence the determination of the $g$-factor, 
for magnetic fields applied perpendicular to the easy plane. 
In this case the frequency-field relationship is no longer linear; 
the applied and internal fields add in quadrature giving rise to a 
finite frequency intercept for the resonance condition at 
zero applied field~\cite{turovbook}. 
For conventional three-dimensional antiferromagnets 
the crossover from paramagnetic resonance to antiferromagnetic 
resonance (with finite zero-field intercept) occurs close 
to the magnetic ordering temperature, $T_{\rm c}$~\cite{Abragam}. 
By contrast, in antiferromagnets of reduced dimensionality 
(in which $T_{\rm c}$ is significantly suppressed relative to the 
energy scale of the dominant exchange interaction) a non-linear 
frequency-field relationship can become evident at temperatures 
significantly above those at which long-range order is observed 
(at $T\simeq 2T_{\rm c}$)~\cite{turovbook,Knaflic}. 
Consequently, the ESR spectra of reduced-dimensional magnetic 
materials must be interpreted with caution.
To obtain the most accurate measure of the paramagnetic 
$g$-factor anisotropy, conditions under which the frequency-field 
relationship becomes strongly nonlinear should be avoided. 
In practice, this means not employing temperatures or frequencies 
that are low relative to the energy scales of the exchange anisotropy. 
Therefore, where possible, for the purpose of this paper, 
we have derived the spectroscopic $g$-factor from the 
gradient of linear fits to measurements at 
multiple high ($\gg 10$~GHz) frequencies at 
temperatures of at least twice $T_{\rm c}$. 
We have also investigated the temperature dependence 
to verify minimal influence of the antiferromagnetic order.

\begin{figure}[t]
\centering
\includegraphics[width=8.5cm]{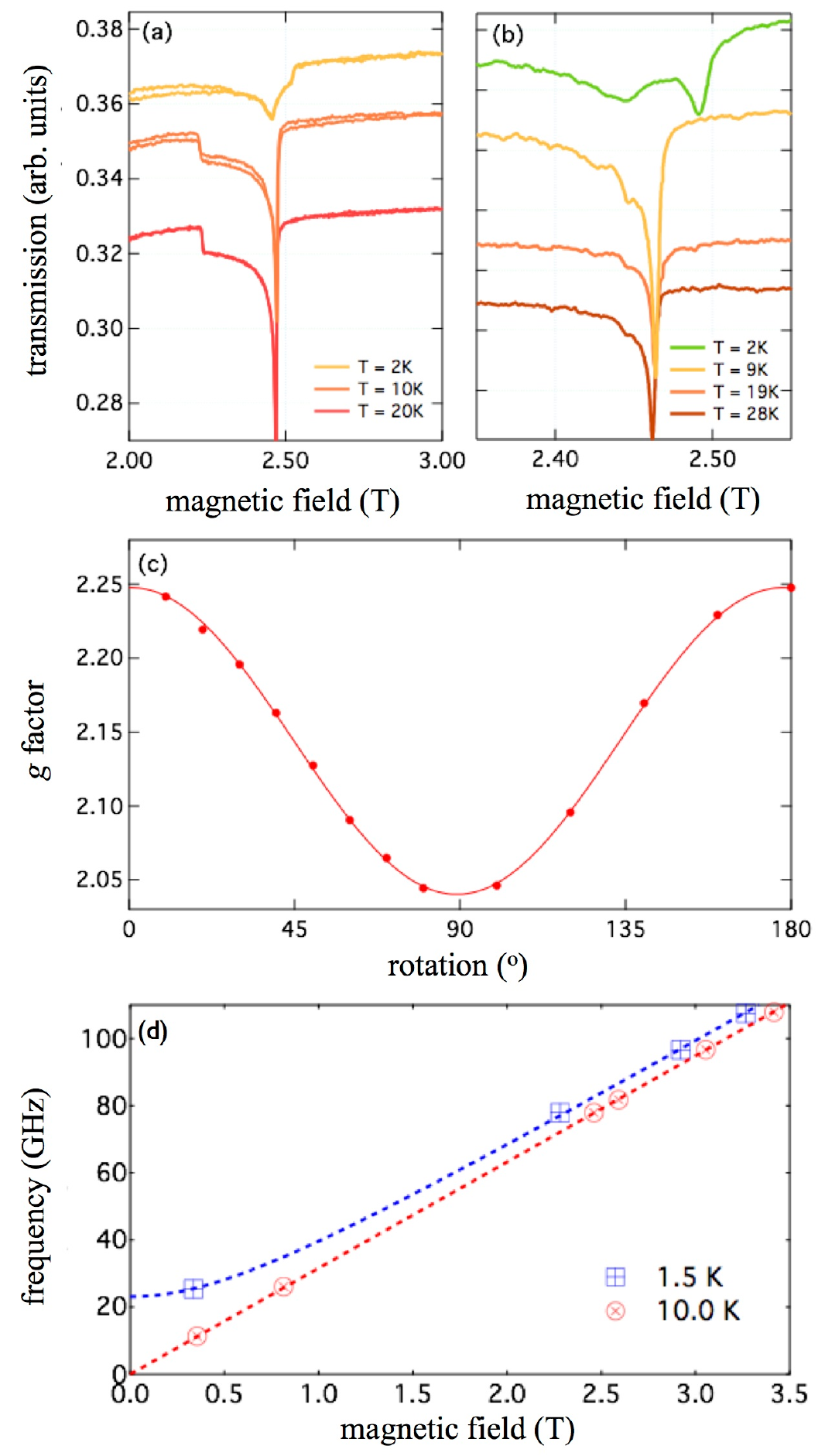}
\sloppypar
\caption{(a)~Powder ESR spectra of  Cu(pyz)$_2$(ClO$_4$)$_2$ ({\bf 4}) 
measured at a frequency of 71~GHz. 
(b)~Similar powder spectra for 
[Cu(4-phpy-O)$_2$(pyz)$_2$](ClO$_4$)$_2$ ({\bf 3}) 
measured at a frequency of 70.3~GHz. 
Above 10~K the $g$-factor anisotropy in both cases is $g_{z}  = 2.25$ 
and $g_{xy} = 2.04$. 
(c)~The $\sin(2\theta$) angle dependance of the g-factor of 
[Cu(pyO)$_2$(pyz)$_2$](ClO$_4$)$_2$ ({\bf 2}) 
as field is rotated between perpendicular to (0$^{\circ}$) and 
parallel to (90$^{\circ}$) the Cu--pyz planes, 
measured at a temperature of 20~K. 
Again the $g$-factor anisotropy is $g_{z}  = 2.25$ and 
$g_{xy} = 2.04$. 
(d)~The frequency -- magnetic field scaling of the ESR 
line measured on single crystal  
[Cu(HF$_2$)(pyz)$_2$]ClO$_4$ ({\bf 1}) with the field oriented 
perpendicular to the planes.}
\label{eprfig}
\end{figure} 

Figure~\ref{eprfig} contains examples of 
ESR measurements on the four materials studied in this paper. 
Figure~\ref{eprfig}(a) shows the powder spectra of {\bf 4} 
measured at a frequency of 71~GHz, while (b) 
shows similar spectra for {\bf 3} measured at a frequency of 70.3~GHz. 
For the latter, a smaller field range is shown to 
emphasize the small shift of the hard axis ESR line in the vicinity 
of antiferromagnetic order at 2~K.  
At temperatures above 10~K the observed asymmetric 
ESR lines in both cases are consistent with a uniaxial 
$g$-factor anisotropy corresponding to 
$g_{z}  = 2.25(5)$ and $g_{xy} = 2.04(3)$. 
Note, these are the $g$-factors obtained with the measurement 
field applied perpendicular and parallel to the 
Cu--pyz planes respectively. 
At a temperature of 2~K the ESR contribution from magnetic 
fields oriented along the easy axis ($g_{z}$) are significantly 
broadened, in agreement with single-crystal measurements 
from other members of this family at temperatures 
approaching that of antiferromagnetic order. 
Figure~\ref{eprfig}(c) shows the angle-dependence of 
the measured $g$-factor in single crystals of {\bf 2} 
as field is rotated between perpendicular to 
(0$^{\circ}$) and parallel to (90$^{\circ}$) the Cu--pyz planes. 
Again the $g-$factor anisotropy is $g_{z}  = 2.25(5)$ 
and $g_{xy} = 2.04(3)$. 
Figure~\ref{eprfig}(d) shows the frequency -- magnetic field 
scaling of the ESR line measured on a single crystal of {\bf 1} 
with the field oriented perpendicular to the planes. 
At a temperature of 10~K the linear frequency-field scaling (with the 0,0 intercept) 
corresponds to a $g$-factor of 2.26(2). 
The non-linear frequency-field scaling observed at a temperature of 
1.5~K (which is below the ordering temperature of 1.91~K) is well 
reproduced by adding a zero magnetic-field frequency offset 
(23~GHz) to the linear frequency-field scaling 
(given by the $g$-factor evolution at high temperatures) 
in quadrature. 
The frequency-field scaling is consistent with evolution to 
antiferromagnetic resonance at low temperature indicating 
an $XY$-type spin exchange anisotropy of a few percent~\cite{Knaflic,Dolinsek}. 

The $g$-factors deduced from all of the ESR experiments
on compounds {\bf 1} to {\bf 4} are reproduced in Table~\ref{magtab};
to within experimental errors, all compounds exibit the same 
range of $g$-factors associated with anisotropy due to
crystal-field effects~\cite{disclaimer}.
The material-independence
of the $g$-factor values shows that the microscopic environment
of the Cu ions is very similar in all four compounds, as
already suggested by the structural data discussed in the previous section.

\begin{figure}[h]
\centering
\includegraphics[width=8.6cm]{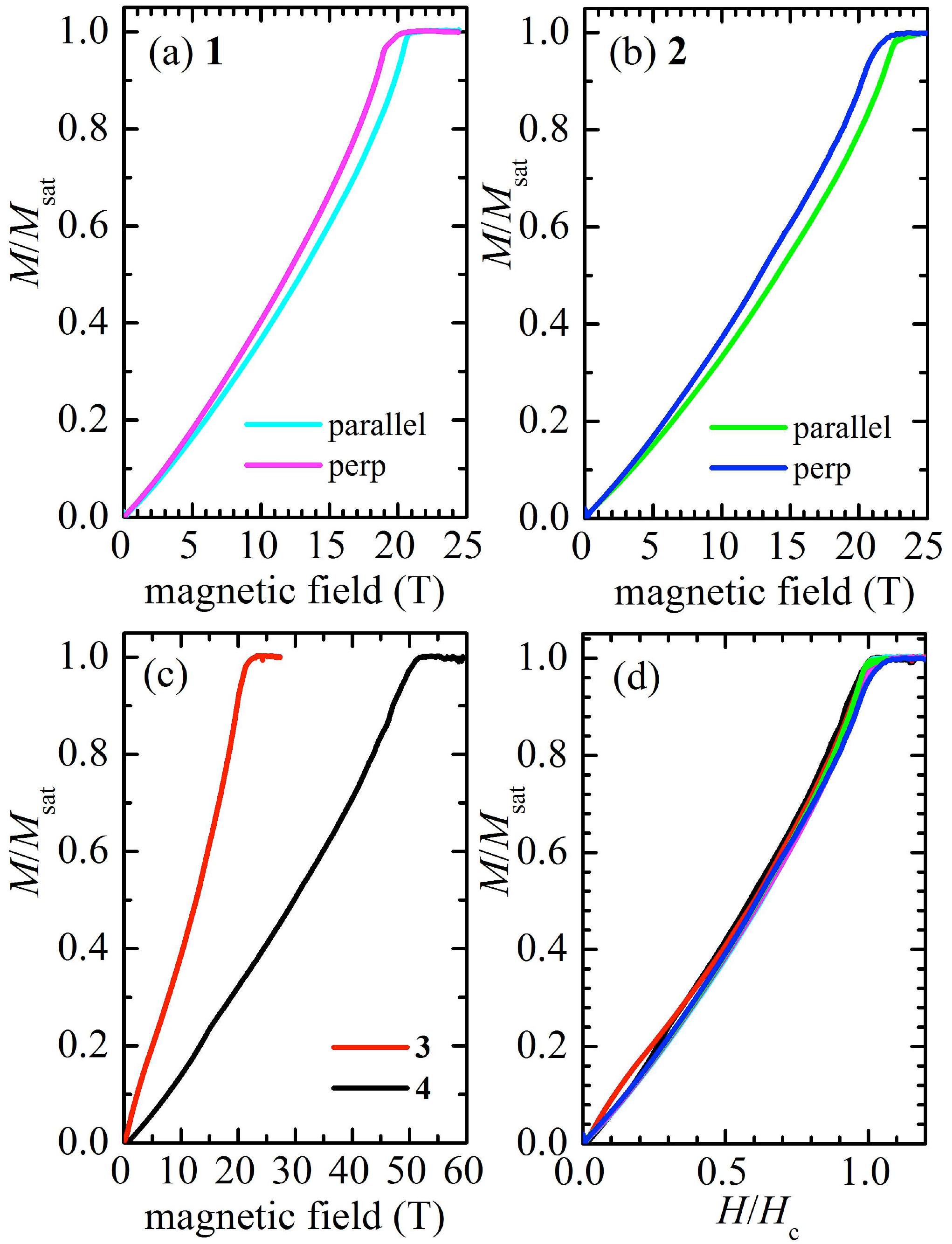}
\sloppypar
\caption{Pulsed-field magnetization versus field. 
Data shown are recorded in increasing fields and at a 
temperature of 0.60~K. (a)~and (b)~show single-crystal data 
for [Cu(HF$_2$)(pyz)$_2$]ClO$_4$ ({\bf 1}) and  
[Cu(pyO)$_2$(pyz)$_2$](ClO$_4$)$_2$ ({\bf 2}), 
respectively, with the field applied parallel and 
perpendicular to the copper-pyrazine planes. 
(c)~Polycrystalline data for  
[Cu(4-phpy-O)$_2$(pyz)$_2$](ClO$_4$)$_2$ ({\bf 3}) and 
Cu(pyz)$_2$(ClO$_4$)$_2$ ({\bf 4}). 
(d)~All of the data versus $H/H_{\rm c}$, where 
$H_{\rm c}$ is the saturation field.} 
\label{magfig}
\end{figure}

\section{Intralayer magnetic exchange energies}
\subsection{Pulsed magnetic field measurements}
Figure~\ref{magfig} shows the magnetization of the 
four compounds measured in pulsed magnetic fields at 
temperatures close to 0.60~K. 
Pulsed-field data have been previously reported 
for {\bf 1}, {\bf 2}~[\onlinecite{njp}] and {\bf 4}~[\onlinecite{turnbull99}]. 
Apart from a low-field hump seen in {\bf 3},
possibly due to a small concentration of paramagnetic
impurities, the form of the low-temperature
magnetization in all cases is very similar: a 
gradual concave rise to the saturation field $H_{\rm c}$. 
This general shape is typical of the $S=1/2$ SLHAFM with 
finite interlayer couplings~\cite{njp} as described by the Hamiltonian
\begin{equation}
{\cal H }= J \sum_{{\left\langle i,j\right\rangle}_{\|}}{S_i \cdot S_j} + J_\perp \sum_{{\left\langle i,j\right\rangle}_{\perp}}{S_i \cdot S_j} - g\mu_{\rm B}B \sum_i{S_i^z}
\label{ham1}
\end{equation}
where $J$ represents the strength of the effective nearest-neighbor 
exchange energy within the planes and  $J_\perp$ the magnetic
coupling between the planes~\cite{njp}. 
The first and second terms describe summations over 
unique pairs of nearest neighbors parallel and perpendicular 
to the planes, respectively, and the last term is the Zeeman energy 
associated with a uniform magnetic field $B$. 
In particular, previously reported quantum-Monte-Carlo 
simulations indicate that data such as those in Figure~\ref{magfig} 
are characteristic of systems for which the spatial exchange anisotropy 
$J_\perp/J\ll 1$ and that the curvature of the 
magnetization increases as this ratio decreases~\cite{njp}. 
This concave $M(H)$ curve is a result of the effect of quantum 
fluctuations on these low-spin systems, which act to 
reduce the moment at a particular field from its classical value. 
If the exchange anisotropy were to be reduced and/or 
the size of the spin quantum number increased the 
magnetization will approach its classical linear form~\cite{njp}.
In Ref.~\onlinecite{njp}, the difference in saturation fields in 
single-crystal measurements along different crystallographic 
axes is attributed to the anisotropy of the $g$-factor. 
This is also the case for the data shown in 
Figures~\ref{magfig}(a) and (b): saturation along different crystallographic 
directions in a particular material occurs at the same value 
of the Zeeman energy once the measured, 
anisotropic $g$ values have been included.
At the saturation field, Equation~\ref{ham1} reduces to 
\begin{equation}
g\mu_{\rm B}B_{\rm c}=nJ+n_\perp J_\perp
\label{Bc}
\end{equation}
where $B_{\rm c}=\mu_0H_{\rm c}$, and for each $S=1/2$ spin, 
$n$ is the number of nearest neighbors within the planes 
and $n_\perp$ is the number perpendicular to the planes. 

For highly anisotropic, square-lattice systems, $n=4$ 
and the final term is small enough to be neglected, allowing 
the size of the intralayer exchange energy to be deduced from 
a measurement of $B_{\rm c}$. 
The saturation fields of three of the compounds shown in 
Figure~\ref{magfig} are similar; 
20.2(2)~T ({\bf 1}), 21.9(2)~T ({\bf 2}) and 21.1(2)~T ({\bf 3}). 
[Note these values in each case represent a polycrystalline average:
$B_{\rm c} = (B_{\rm c}^z + 2B_{\rm c}^{xy})/3$.] 
Using the above relation and the relevant $g$-factors from the 
ESR experiments we find that $J=7.2(2)$~K ({\bf 1}), 
7.7(2)~K ({\bf 2}) and 7.5(2)~K ({\bf 3}). 
These values are tabulated in Table~\ref{magtab}.

The size of $J$ for {\bf 1} is comparable to that of 
[Cu(HF$_2$)(pyz)$_2$]$X$ with $X={\rm BF}_4$, but smaller by 
nearly a factor of two than the members of the same family with 
octahedral counterions $X={\rm PF}_6$, SbF$_6$ and AsF$_6$~\cite{njp}. 
In contrast, changing the symmetry of the counterion in 
the [Cu(pyO)$_2$(pyz)$_2$]($X$)$_2$ family does not appear to 
have such a significant effect on the exchange energy: 
the $X={\rm PF}_6$ material has $J=8.1(3)$~K~\cite{godprl}, 
only slightly larger that found for {\bf 2}. 

The similarity in $J$ between compounds {\bf 1}, {\bf 2} and {\bf 3}
is not unexpected, given their comparable copper-pyrazine 
square-lattice configurations. 
What is more surprising is that, as shown in Figure 3(c), {\bf 4} 
requires the application of a much bigger field ($B_{\rm c} = 51.1(2)$~T) 
to fully align the spins, yielding a effective exchange energy of 
$J=18.1(4)$~K, which is in reasonable agreement with 
previous estimates~\cite{woodward1, turnbull99}. 
The square-lattice network in {\bf 4} is not unlike those in the 
other compounds; why, then, should this exchange energy be so different? 
One major structural difference between {\bf 4} 
and the other materials considered here is that in the 
former the ClO$_4$ molecules are coordinated to the 
copper-ions rather than adopting non-coordinated 
positions centrally above the square copper-pyrazine plaquettes. 
Although the ClO$_4$ ligands in {\bf 4} do indeed sit 
in near-central locations on either side of the squares of the 
neighboring planes, this contrast in the way the molecule 
is bonded will likely lead to a difference in the distribution 
of electron density close to the layers for this 
compound as compared to the other three. 

We recall again that the reduction in $J$ observed on changing 
the counterion symmetry in the [Cu(HF$_2$)(pyz)$_2$]$X$ 
family from octahedral to tetrahedral is accompanied by 
a change in the tilt angle of the pyrazines, which could also 
act to redistribute electron density. 
Pyrazine tilting has been correlated with changes in $J$ 
in copper-pyrazine square-lattice materials on several 
occasions~\cite{njp, steele,darriet}, 
but a causal relationship has not been established. 
However, in the context of the current paper,
the X-ray measurements discussed above
showed that the pyrazine tilt angles are similar for all four of 
the compounds that are the focus of this paper.

\renewcommand{\arraystretch}{1.2}
\begin{table*}[t]
\centering
\newcolumntype{C}{ >{\centering\arraybackslash} m{1.0cm} }
\newcolumntype{D}{ >{\centering\arraybackslash} m{1.2cm} }
\begin{tabular}{m{4.9cm}*{1}{D}*{1}{D}*{3}{C}*{1}{D}*{2}{C}*{1}{D}*{1}{D}}
\hline\hline
											& $g$-factor		& $\mu_0H_{\rm c}$		& $J$ 	& $T_{\rm c}$	 & $T_{\rm c}/J$	& $|J_{\perp}/J|$	& $J_{\perp}$	 	& $E_{\rm D}^\perp$	& $\mu_0H_{\rm A}$	& $\Delta\times J$	\\ 
											&				& (T)					& (K)		& (K)			&				&				& (mK)			& (mK)			& (T)				& (mK)			\\ \hline
({\bf 1}) [Cu(HF$_2$)(pyz)$_2$]ClO$_4$ 				& 2.25(5) 2.07(3)	& 20.2(2) 				& 7.2(2) 	& 1.91(1)		& 0.27			& $2\times10^{-3}$	& 14				& 4.7				& 0.08			& 28				\\ \hline
({\bf 2}) [Cu(pyO)$_2$(pyz)$_2$](ClO$_4$)$_2$ 		& 2.25(5) 2.04(3)	& 21.9(2) 				& 7.7(2) 	& 1.70(1)		& 0.22 			& $3\times10^{-4}$	& 2 				& 0.11 			& 0.11			& 39 				\\ \hline
({\bf 3}) [Cu(4-phpy-O)$_2$(pyz)$_2$](ClO$_4$)$_2$ 	& 2.25(5) 2.04(3)	& 21.1(2) 				& 7.5(2) 	& 1.63(1)		& 0.22 			& $3\times10^{-4}$	& 2 				& 0.02			& 0.11			& 39				\\ \hline	
({\bf 4}) Cu(pyz)$_2$(ClO$_4$)$_2$  				& 2.25(5) 2.04(3)	& 51.1(2) 				& 18.1(4) 	& 4.21(1)		& 0.23 			& $5\times10^{-4}$ 	& 9 				& 2.6				& 0.28			& 100			\\ \hline \hline
\end{tabular}
\caption{Magnetic parameters. The anisotropic $g$-factor 
is derived from frequency-dependent electron-spin-resonance 
measurements performed at 24~K ({\bf 1}), 20~K({\bf 2}), 
28~K ({\bf 3}), and 19~K ({\bf 4}). 
$\mu_0H_{\rm c}$ and $J$ are, respectively, the powder-averaged 
saturation magnetic field and the effective nearest-neighbor 
exchange energy deduced from pulsed-field-magnetization 
measurements performed at 600~mK. 
The antiferromagnetic ordering temperature ($T_{\rm c}$) is 
established via muon-spin relaxation studies. 
The  $T_{\rm c}/J$ ratios are used to estimate the upper 
bounds of the interplane coupling ($J_{\perp}$) and 
spatial exchange anisotropy ($|J_{\perp}/J|$) in the 
Heisenberg model~\cite{yasuda}. 
Dipole-field calculations are used to estimate the energy 
scale of the interplane dipolar interaction $E_{\rm D}^\perp$ 
for an ordered moment of $1~\mu_{\rm B}$. 
The anisotropy field $\mu_0H_{\rm A}$ is estimated from the 
position of kinks in the magnetization data at low temperatures. 
The spin-exchange anisotropy, $\Delta$, is given by the ratio 
$H_{\rm A}/H_{\rm c}$ and the anisotropy energy scale is 
parameterized by the product of $J$ and $\Delta$.
}
\label{magtab}
\end{table*}

\subsection{Density-functional-theory calculations}

The nature of the super-exchange interactions in this class of 
material has been the subject of attention in the past, with early 
experimental and theoretical studies at odds as to the 
relevant importance of the 
$\sigma$-orbitals~\cite{mohri99} over the pyrazine 
$\pi$-orbitals~\cite{darriet,richardson77}.
More recent theoretical studies have made use of density-functional 
theory (DFT) and, in particular, compound {\bf 4} has been 
investigated using first-principles calculations by Vela {\it et al.}~\cite{vela13}. 
According to their calculations a disparity of about 
30\% (21.0~K vs 29.6~K) is present at 15~K between the 
exchange strengths through the two crystallographically distinct 
pyrazine molecules in this material~\cite{vela2}.
By contrast, using the 165~K structure their calculations 
show only a very small difference in these exchange energies. 
Vela {\it et al.}~\cite{vela13}  rule out the tilting of the pyrazines as a 
contributory factor in the low-temperature disparity and instead 
attribute it to a combination of three effects, which are, in 
increasing order of importance: (i)~hydrogen bonding between 
the O atoms of the perchlorates and the H atoms of the pyrazines; 
(ii)~a shear-like distortion of the pyrazine rings; and (iii)~the orientation 
of the ClO$_4^-$ ligands. 
The authors of Ref.~\onlinecite{vela13} suggest that the role of 
ClO$_4^-$ molecules is to increase the spin-density 
along the primary exchange pathway leading to an enhanced 
interaction strength. By extension, for the other samples 
considered here, where the ClO$_4^-$ counterions occupy 
voids between the layers, this enhancement is not expected to 
occur, which could explain the large disparity in the 
in-plane exchange energies.

We have performed DFT studies on all four of the compounds 
using the low-temperature structures shown in Figure~\ref{strucfig}. 
In each case, two distinct exchange strengths were identified 
along the Cu-pyrazine linkages within the unit cell. 
These are found to be 4.3 and 10.2~K ({\bf 1}), 4.7 and 8.1~K ({\bf 2}), 
14.5 and 17.0~K ({\bf 3}), and 22.8 and 21.3~K ({\bf 4}). 
We note that the disparity in the two exchange strengths 
for {\bf 4} is smaller than that found in Ref.~\cite{vela13}. 
This is probably because the authors of that work used a 10~K, 
rather than a 100~K, structure; the disparity develops on cooling. 
If the predictions are correct it would imply that these materials 
correspond to a rectangular rather than a square-lattice model.  
It is not presently possible to experimentally verify such 
disparities in the exchange parameters from temperature-dependent 
magnetic susceptibility data owing a lack of analytical fitting expressions 
that can adequately discriminate between square and rectangular
models of antiferromagnetism. 
We can, however, compare the theoretical results with the 
exchange energies listed in Table~\ref{magtab}, 
obtained by applying the SLHAFM model to the 
low-temperature pulsed-field magnetization measurements, 
by calculating an average or effective intralayer 
exchange strength from the DFT calculations. 
These are 7.3~K ({\bf 1}), 6.4~K ({\bf 2}), 15.8~K ({\bf 3}), and 22.1~K ({\bf 4}). 
The correspondence between experiment and calculation 
is good for {\bf 1} and {\bf 4}, reasonable for {\bf 2}, but poor for {\bf 3}. 
The disappointing result for {\bf 3} may be a consequence of the
difficulties of treating the 3d-4s mixing in the perchlorate ligand, 
which is a relatively poor donor~\cite{EinM}. 
\subsection{Next nearest-neighbor interactions}
Finally in this section, we point out that other exchange
pathways in the Cu--pyz planes are also possible. 
A recent neutron-scattering study of {\bf 4}~\cite{tsyrulinprl} 
analysed the intralayer spin-wave spectrum in this material 
using the isotropic $J_1-J_2$ model, which has two 
intralayer exchange terms in the Hamiltonian:
$J_1 \sum{S_i \cdot S_j} + J_2 \sum{S_i \cdot S_k}$, 
where $J_1$ is nearest-neighbor exchange energy along the 
Cu--pyz--Cu exchange paths, $J_2$ is an additional 
next-nearest-neighbor exchange across the diagonal of the 
copper-pyrazine squares and the summations are over the 
unique spin pairs associated with these exchange pathways. 
If both $J_1$ and $J_2$ are antiferromagnetic, then they will act to 
frustrate one another and the effective nearest-neighbor 
exchange energy of Equation~\ref{ham1} will be given by $J = J_1 - J_2$. 
Theory predicts that if the $J_2$ is a significant fraction of 
$J_1$ the ground state of this model is no longer the 
simple N\`{e}el state, but transforms first into a disordered 
spin-liquid phase and, at higher values of $J_2$, an ordered collinear state~\cite{chandra88,sushkov01,viana07}.

However, data described in the next section
show that all the materials considered here exhibit a relatively simple
phase diagram, with a low-temperature ordered phase. 
This is in accord with expectations that  $J_2$ is small due 
to the lack of an effective exchange pathway across the diagonal. 
The authors of Ref.~\cite{tsyrulinprl} estimated $J_2 \simeq 0.02J_1$ 
from their spin-wave data on {\bf 4}. 
A later neutron spectroscopy study~\cite{tsyrulinprb} expanded this 
analysis to conclude a small $XY$ anisotropy in $J_1$, 
which will be discussed in more detail later. 

To summarize this Section, the intralayer exchange energies
in compounds {\bf 1}-{\bf 3} are probably chiefly determined
by nearest-neighbour interactions via Cu--pyz--Cu
exchange pathways, yielding a dominant $J\approx 7.5$~K
in all cases. The almost identical values of the intralayer exchange
energies in these three compounds is in accord with the earlier sections
of the current paper, in which both the structure of the
Cu--pyz planes and the local crystal-field environment
of Cu are shown to be very similar in {\bf 1}, {\bf 2} and {\bf 3}.

\section{Interactions determining the magnetic ordering temperatures}
\subsection{Muon spin-rotation measurements}
Reduced-dimensionality magnetic systems accommodate 
short-range spin correlations that begin to build up at temperatures 
higher than the transition temperature, affecting the thermodynamic 
properties of the system~\cite{Knaflic,sengupta03}. 
This leads to the observation of a broad maximum in both the 
magnetic susceptibility and the heat capacity at temperatures of 
the order of the primary exchange energy~\cite{njp}. 
At lower temperatures additional interactions force the system into a state 
of long-range order, but the associated entropy change at the 
ordering temperature can be rather small, 
because the spins are already highly correlated. 
The result is that the feature by which the transition 
can be identified is often masked by the much larger 
hump resulting from the spin correlations~\cite{sengupta03}. 
Local probes, such as muon-spin rotation ($\mu^+$SR), 
do not suffer from the same drawbacks as thermodynamic techniques 
because they are sensitive to the static, internal magnetic 
field which only develops once three-dimensional 
long-range order sets in~\cite{njp,Blundell99}. 
Thus $\mu^+$SR can be used to determine the 
critical temperature in highly anisotropic materials.

\begin{figure}[t]
\centering
\includegraphics[width=8.0cm]{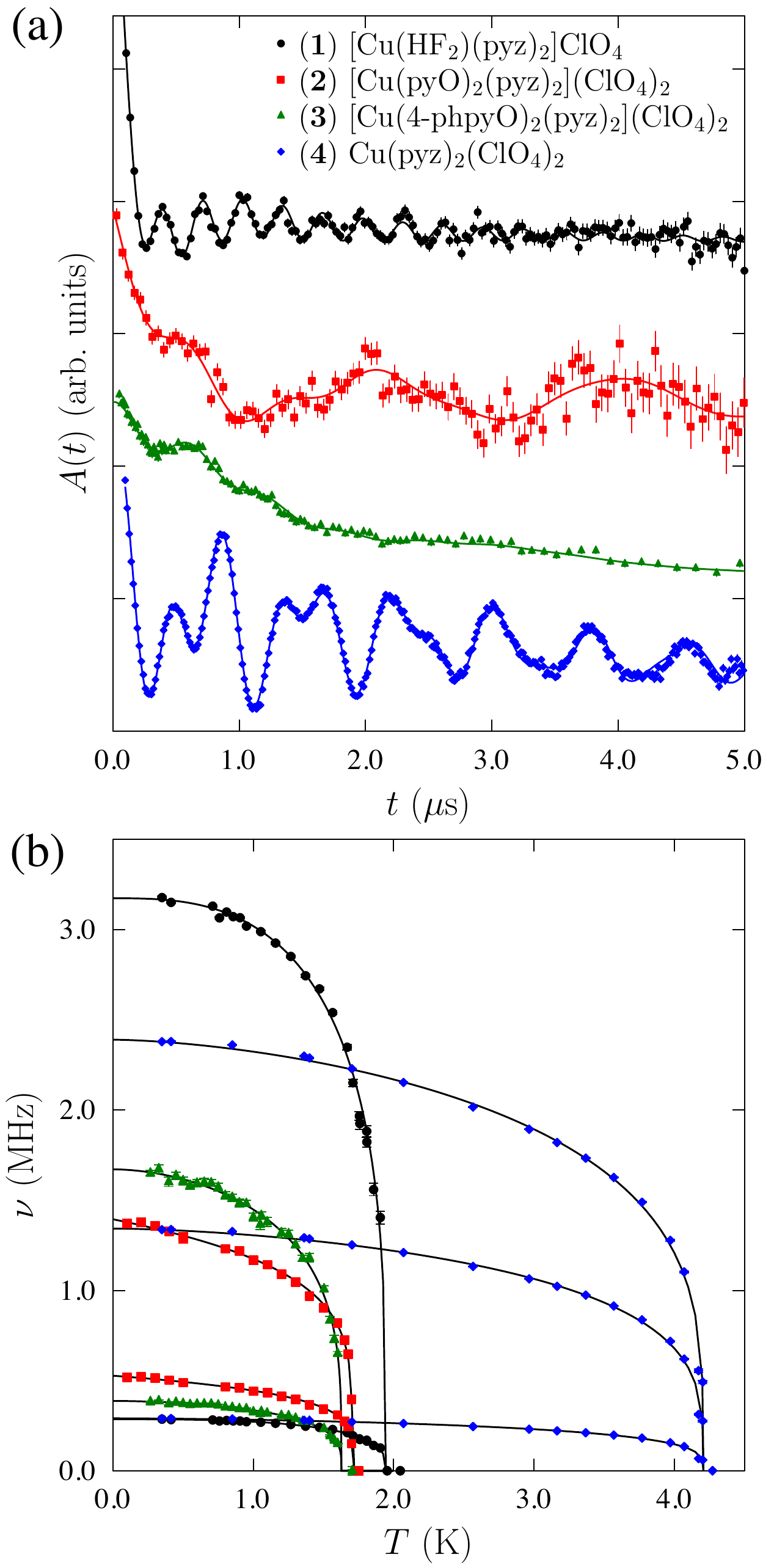}
\sloppypar
\caption{(color online). (a)~Muon asymmetry observed at 
0.41~K ({\bf 1}), 0.1~K ({\bf 2}), 0.26~K ({\bf 3}), 
and 0.34~K ({\bf 4}); points are data, whilst lines represent
the fits described in the text.
The data for {\bf 1} and {\bf 4} are taken from~\cite{steele} 
and~\cite{lancasterprb07}. (b)~Temperature evolution of the oscillation frequencies $\nu$. 
The solid lines are fits to the functional form  
$\nu_i(T)= \nu_i(0)[1-(T/T_{\rm c})^\alpha]^\beta$ as described in the text.}
\label{muonfig}
\end{figure}

In a $\mu^{+}$SR experiment spin-polarized positive muons are 
implanted in a target sample, where the muon usually occupies an
interstitial position in the crystal~\cite{Blundell99}. 
In such an experiment, the time evolution of the muon-spin polarization
is observed, the behavior of which depends on the local magnetic 
field at the muon site. 
Each muon decays, with an average lifetime of 2.2~$\mu$s, 
into two neutrinos and a positron, 
the latter particle being emitted preferentially along the instantaneous 
direction of the muon spin. 
Recording the time dependence of the positron emission directions 
therefore allows a determination of the spin-polarization of the 
ensemble of muons as a function of time. 
Histograms $N_{\mathrm{F}}(t)$ and $N_{\mathrm{B}}(t)$ record the 
number of positrons detected in detectors placed forward (F) and backward 
(B) of the initial muon polarization direction. 
The quantity of interest is the positron-decay asymmetry function, 
defined as
\begin{equation}
A(t)=\frac{N_{\mathrm{F}}(t)-\gamma N_{\mathrm{B}}(t)}
{N_{\mathrm{F}}(t)+\gamma N_{\mathrm{B}}(t)} \, ,
\end{equation}
where $\gamma$ is an experimental calibration constant. 
$A(t)$ is proportional to the spin polarization of the muon ensemble.

Example spectra are shown in Fig.~\ref{muonfig}(a). 
Below a characteristic temperature, oscillations in the 
asymmetry are observed in all samples. 
This is because the local field causes a coherent precession of the 
spins of those muons for which a component of their spin polarization 
lies perpendicular to this local field~\cite{Blundell99}. 
Thus the observation of oscillations provides clear evidence 
for long-range magnetic order throughout the bulk of the sample.

\renewcommand{\arraystretch}{1.1}
\begin{table}[t]
\centering
\newcolumntype{C}{m{0.8cm} }
\newcolumntype{D}{ >{\centering\arraybackslash} m{1.1cm} }
\begin{tabular}{C*{7}{D}}
\hline \hline
		& $\nu_1(0)$ (MHz)		& $\nu_2(0)$ (MHz)		& $\nu_3(0)$ (MHz)		& $\alpha$	& $\beta$		& $T_{\rm c}$ ~~(K) 	\\
\hline
({\bf 1})	& 3.2(1)				& 6.4(1)				& --					& 2.6(3)		& 0.25(2)		& 1.91(1)			\\
({\bf 2})	& 1.39(1)				& 0.53(1)				& --					& 0.8(2)		& 0.18(1)		& 1.70(1)			\\
({\bf 3})	& 1.67(2)				& 0.39(1)				& --					& 1.8(2)		& 0.27(2)		& 1.63(1)			\\
({\bf 4})	& 2.38(3)				& 1.33(2)				& 0.29(2)				& 1.8(3) 		& 0.28(2)		& 4.21(1)			\\
\hline \hline
\end{tabular}
\caption{Parameters extracted from fits to the muon-precession 
frequencies shown in Figure~\ref{muonfig}(b) for 
({\bf 1}) [Cu(HF$_2$)(pyz)$_2$]ClO$_4$, 
({\bf 2}) [Cu(pyO)$_2$(pyz)$_2$](ClO$_4$)$_2$, 
({\bf 3}) [Cu(4-phpyO)$_2$(pyz)$_2$](ClO$_4$)$_2$, 
and ({\bf 4}) Cu(pyz)$_2$(ClO$_4$)$_2$. 
The parameters for ({\bf 1}) and ({\bf 4}) are taken 
from~\cite{steele} and~\cite{lancasterprb07}. 
}
\label{muontab}
\end{table}

For {\bf 2} at $T\leq1.7$~K the spectra were fitted to 
\begin{eqnarray}
A(t)=A_1 e^{-\lambda_1 t} \cos(2 \pi \nu_1 t + \phi_1) + \nonumber \\ A_2 e^{-\lambda_2 t} \cos(2 \pi \nu_2 t + \phi_2) + \nonumber \\ A_3 e^{-\lambda_3 t} + A_{b},
\label{muonafit}
\end{eqnarray}
where the first two terms account for muons whose spins precess 
coherently in a quasistatic magnetic field, the third term is 
due to muons that depolarize too rapidly for oscillations 
to be observed and the final term accounts for the muon-spin projection 
parallel to the local magnetic field and muons that stop 
in the sample holder or cryostat tail~\cite{Blundell99}. 
Non-zero phase angles $\phi_i$ were found to be necessary;
they did not correlate with $\nu_i(T)$. 
During the fitting process we fixed $\phi_1=-29^\circ$ and 
$\phi_2=40^\circ$. 
A zeroth-order Bessel function of the first kind 
(indicative of spin-density wave order~\cite{Le1993}) was found 
to give a worse fit to the experimental spectra. 
The ratios $A_1/A_2$ and $\nu_2/\nu_1$ 
were fixed to 0.864 and 0.378, respectively. 
Close to the transition it was necessary to also 
fix $A_2=2.91$\% (and thus $A_1=2.51$\%). 
The asymmetry spectra for {\bf 3} at $T\leq1.6$~K 
were fitted to the same functional form but with 
$A_b=A_0\exp(-\Gamma t)$ instead of a constant. 
The fits to the low-temperature spectra for {\bf 1} and 
{\bf 4} also use Equation~\ref{muonafit} (with the 
addition of a third precession frequency for {\bf 4}) 
and have been previously described elsewhere~\cite{lancasterprb07,steele}. 

The temperature-dependences of the fitted precession 
frequencies are shown in Figure~\ref{muonfig}(b). 
The oscillation frequencies represent an effective order parameter~\cite{Blundell99} 
and were fitted to the phenomenological expression 
$\nu(T)=\nu(0)[1-(T/T_{\rm c})^\alpha]^\beta$, yielding the parameters 
listed in Table~\ref{muontab}. 
The extracted values of the $\beta$ critical parameter 
are typical of reduced-dimensionality magnetic 
interactions~\cite{steele}. 
The critical temperatures are also tabulated in 
Table~\ref{magtab}, and are seen to decrease slowly 
across the family {\bf 1}, {\bf 2} and {\bf 3} as 
the interlayer separation is increased. 
Compound {\bf 4} has a significantly 
higher $T_{\rm c}$ than {\bf 1}-{\bf 3},
the reasons for which will be discussed below.

\subsection{Interlayer exchange interactions}
Ideal two-dimensional Heisenberg magnets do not exhibit 
long-range order at temperatures above zero kelvin~\cite{mermin}. 
Real systems, however, have additional interactions that act to 
promote magnetic ordering. 
In particular, for the quasi-two-dimensional model 
described by Equation~\ref{ham1}, a finite spatial exchange 
anisotropy ($|J_\perp/J|$) enhances $T_{\rm c}$ according 
to the empirically-derived relation~\cite{yasuda}
\begin{equation}
T_{\rm c}/J = 2.30/(2.43-\ln |J_\perp/J|),
\label{yasudaeqn}
\end{equation}
which has been shown to be valid in the absence of strong 
quantum-critical fluctuations~\cite{hastings06, praz06}. 
The $T_{\rm c}/J$ ratios for the compounds considered here 
are shown in Table~\ref{magtab} and are seen to be similar 
across the family, but with compound {\bf 1} having a slightly 
higher value than the others. 
Also shown are the estimates of $|J_\perp/J|$ and $|J_\perp|$ obtained 
by applying the above formula. 
According to this calculation all four compounds can 
be considered to be highly two-dimensional with spatial-exchange 
anisotropies $\le 10^{-3}$ (although it should be noted 
that this requires extrapolation of Equation~\ref{yasudaeqn} 
somewhat beyond the range for which it was originally 
derived~\cite{yasuda}). 
By this estimate {\bf 2} and {\bf 3} are among the closest 
approximations to an ideal SLHAFM of any copper-pyrazine 
system realised so far, having a spatial exchange ratio (0.03\%) 
slightly lower than that of the related compound 
[Cu(2-pyridone)$_2$(pyz)$_2$](ClO$_4$)$_2$ (0.05\%)~\cite{selmani10}, 
but not as low as the inorganic SLHAFM 
Sr$_2$CuO$_2$Cl$_2$ (0.003\%)~\cite{lancasterprb07}, 
or the $S=1/2$ coordination polymer 
Ag(pyz)$_2$(S$_2$O$_8$) (0.0002\%)~\cite{mansonag}. 

The estimated interlayer couplings are similar for 
{\bf 1}  and {\bf 4}, which have comparable interlayer spacings 
(see Table~\ref{structab}); this implies that the 
difference in ordering temperatures in these two materials arises 
predominantly from the variation in their intralayer exchange 
strength described in the previous section. 

As the layer separation increases by a factor $\approx 2.5$
between compounds {\bf 1} and {\bf 3},
there is an accompanying sevenfold reduction in the 
interlayer coupling strength. 
If we naively attribute the drop in interlayer coupling to 
a power-law dependence on interlayer separation 
($J_\perp \propto R^{-p}$), we find that the exponent $p\simeq 2.1$. 
In several systems superexchange energies have 
empirically been found to vary with interionic spacing 
with the exponent $p\approx 10$ (see e.g.~\cite{bloch66,hutchings68,rogers72}), 
a result which has been interpreted to be a consequence 
of the role of $\sigma$-bonding in the superexchange 
process in these materials~\cite{johnson74}. 
Magnetic exchange is contingent on the degree of overlap 
between neighboring electronic orbitals and its spatial variation 
will depend upon the nature of the exchange pathways. 
Unlike the Cu compounds that are the subject of this paper,
none of the materials described in 
Refs.~\onlinecite{bloch66,hutchings68,rogers72}
involve exchange mediated through aromatic molecules. 
Therefore the exchange interactions may 
not necessarily be expected to have the same distance dependence. 
Nevertheless, as the exponent $p \approx 2.1$ for compounds
{\bf 1}-{\bf 3} is much less than that 
expected for superexchange ($p\approx 10$; see above), in the following section,
we consider other 
types of interaction that might 
be present and that could drive magnetic order.

The interlayer exchange energies in Table~\ref{magtab}
were derived using the Hamiltonian in Equation~\ref{ham1}. 
Since, as discussed below, other interactions may contribute to 
the magnetic order, the values of the spatial exchange anisotropy quoted in 
Table~\ref{magtab} should be considered as upper bounds.

\subsection{Dipolar interactions}
Given the small energy scales in these molecular systems, 
it is important to assess to what extent 
symmetry-breaking dipolar interactions contribute to the 
magnetic anisotropy and the propensity of the system to 
undergo magnetic ordering. 
This issue provoked
early theoretical attention by Luttinger and Tisza~\cite{Luttinger1946} 
as well as more recent experimental work in a rare-earth system~\cite{Kraemer2012}. 
In principle, long-range order in quasi-2D Heisenberg antiferromagnets 
can be brought about by the dipolar couplings between the layers. 
The dipole interaction has a $R^{-3}$ distance
dependence~\cite{kotler14,kurmooferro,kurmoo99}, 
which is similar to the estimate determined above by comparing the 
upper bounds of the interlayer exchange couplings in {\bf 1} and {\bf 3}. 

We have performed dipole-field calculations given 
a certain assumed collinear antiferromagnetic structure 
inspired by the structure suggested for {\bf 4} on the basis 
of neutron-diffraction experiments~\cite{tsyrulinprb}. 
The calculations leave the direction of each individual 
moment variable. and hence allow the dipolar 
anisotropy to be determined for the assumed magnetic structure;
results are shown in Table~\ref{magtab}.
The interlayer dipolar coupling $E_{\rm D}^\perp$ decreases 
rapidly with interlayer separation 
and is largest for {\bf 1}. 
For {\bf 1} and {\bf 4} $E_{\rm D}^\perp$ is found to be 
similar to the magnitude of interlayer coupling necessary to 
account for the measured $T_{\rm c}/J$ ratio, suggesting that 
dipolar interactions could play a significant role in driving 
the magnetic ordering in these compounds. 

We point out that our dipolar calculations assume an ordered moment 
size of 1$\mu_{\rm B}$. 
Quantum fluctuations are expected to reduce the moment in 
reduced-dimensionality magnets and an ordered moment of 
0.62$\mu_{\rm B}$ is expected for the ideal 2D 
SLHAFM at zero temperature~\cite{manousakis}. 
In such circumstances,
the quoted energy scales of the dipolar interactions must hence be 
reduced by a factor of up to 0.38.
However, the ordered moment of Cu(pyz)$_2$(ClO$_4$)$_2$
has been measured to be as low as $0.47(5)\mu_{\rm B}$
(see Ref.~\onlinecite{tsyrulinprl}), so the dipolar interactions
could potentially be reduced even further.
 
\subsection{Spin exchange anisotropy}
The preceding analyses attribute the finite value of 
$T_{\rm c}$ entirely to some kind of coupling between the 
copper-pyrazine planes.  
However, it has been suggested that the magnetic order in these types of 
material could be driven by a combination of interlayer coupling and 
spin-exchange anisotropy within the layers (a departure 
from a Heisenberg-type interactions)~\cite{fan}. 
The $g$-factor anisotropy determined above shows that the 
spin-orbit coupling that gives rise to spin-exchange anisotropy 
is present in the materials considered here. 
The Hamiltonian relevant to these systems is consequently amended to
\begin{eqnarray}
{\cal H }= J_1 \sum_{{\left\langle i,j\right\rangle}_{\|}}[S_i^xS_j^x+S_i^yS_j^y+(1-\Delta)S_i^zS_j^z] \nonumber \\
+ J_2 \sum_{{\left\langle i,k\right\rangle}_{\|}}{S_i \cdot S_k} 
+ J_\perp \sum_{{\left\langle i,j\right\rangle}_{\perp}}{S_i \cdot S_j} \nonumber \\
 - g\mu_{\rm B}B \sum_i{S_i^z}
\label{hamfull}
\end{eqnarray}
where we have included both nearest-neighbor ($J_1$) and 
next-nearest-neighbor ($J_2$) intralayer interactions, as 
well as the interlayer coupling ($J_\perp$). 
$\Delta$ is the spin-exchange anisotropy parameter and is expected to 
be small and positive ($XY$-like) for quasi-two-dimensional 
$S=1/2$ systems. 
Ignoring the $J_2$ and $J_\perp$ terms, quantum-Monte-Carlo 
simulations predict a Berezinskii-Kosterlitz-Thouless-type 
transition to three-dimensional magnetic order driven by even 
very small levels of spin exchange anisotropy~\cite{cuccoliprb03,dingprl92} 
and obtain $T_{\rm c}/J$ ratios similar to those measured here 
arising from $\Delta\sim 10^{-2}$. 
Reference~\onlinecite{fan}, again neglecting any effects of $J_2$, 
found evidence for spin-exchange anisotropy in low-field magnetization 
and low-temperature magnetic susceptibility measurements of 
several quasi-two-dimensional copper-pyrazine polymeric magnets, 
including compound {\bf 4}. 
The observation of a change in slope, or kink, in 
$M(H)$ obtained with the field applied parallel to the planes 
allowed the authors to conclude $\Delta = 5\times 10^{-3}$  
for {\bf 4}~[\onlinecite{LaterESR}]. 
Using neutron-spectroscopy measurements on the 
aforementioned compound, some of the same authors later 
determined values of $J_2\simeq0.02J_1=0.4$~K and $\Delta=2\times10^{-3}$~[\onlinecite{tsyrulinprb}]. 
ESR~\cite{cizmar10} and heat capacity~\cite{kohama10} 
data were used to estimate $\Delta = 3\times10^{-3}$ and 
$7\times10^{-3}$ for the PF$_6^-$ counterion analogues 
of {\bf 1} and {\bf 2}, respectively. 

\begin{figure}[t]
\centering
\includegraphics[width=8.5cm]{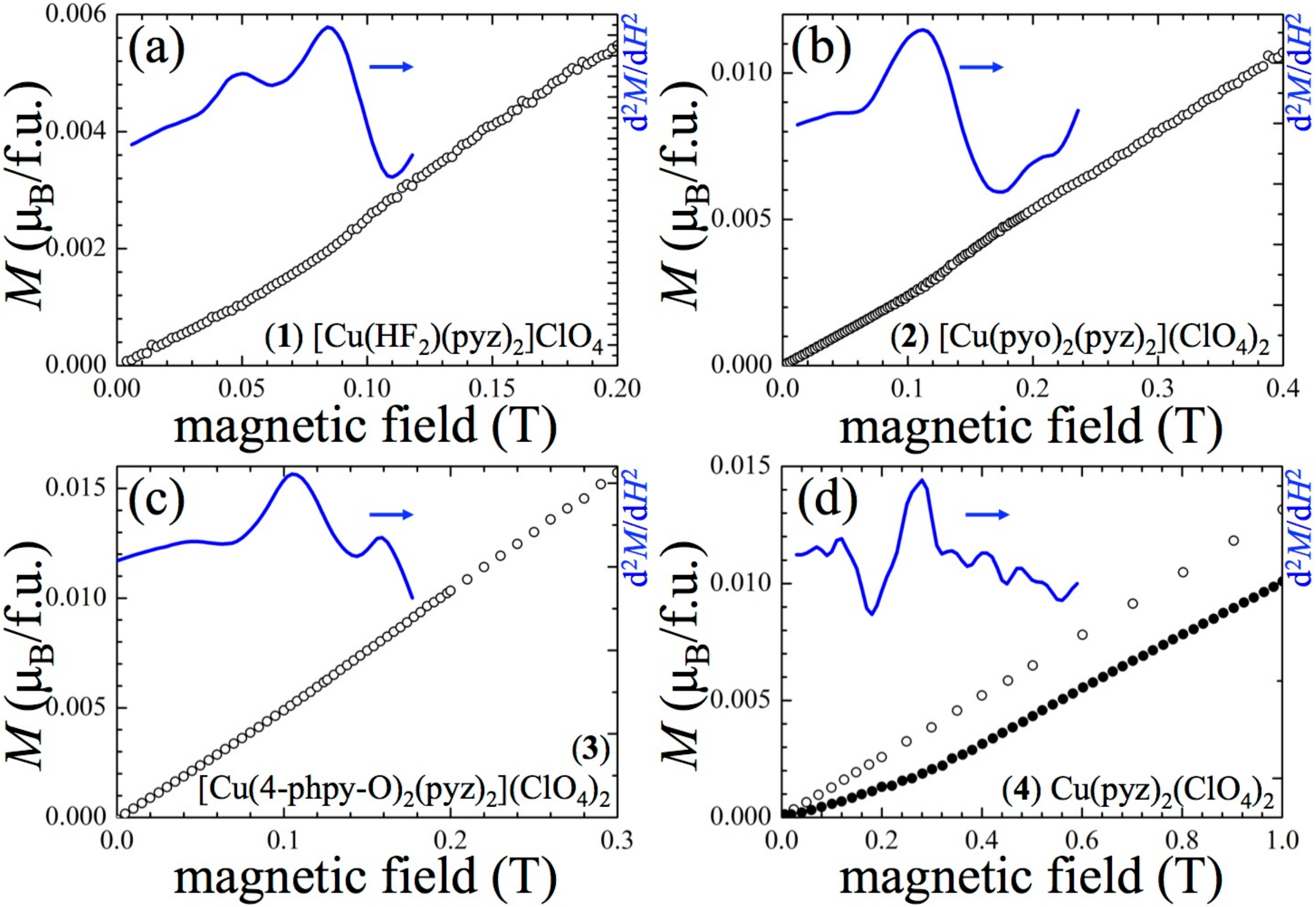}
\includegraphics[width=8.5cm]{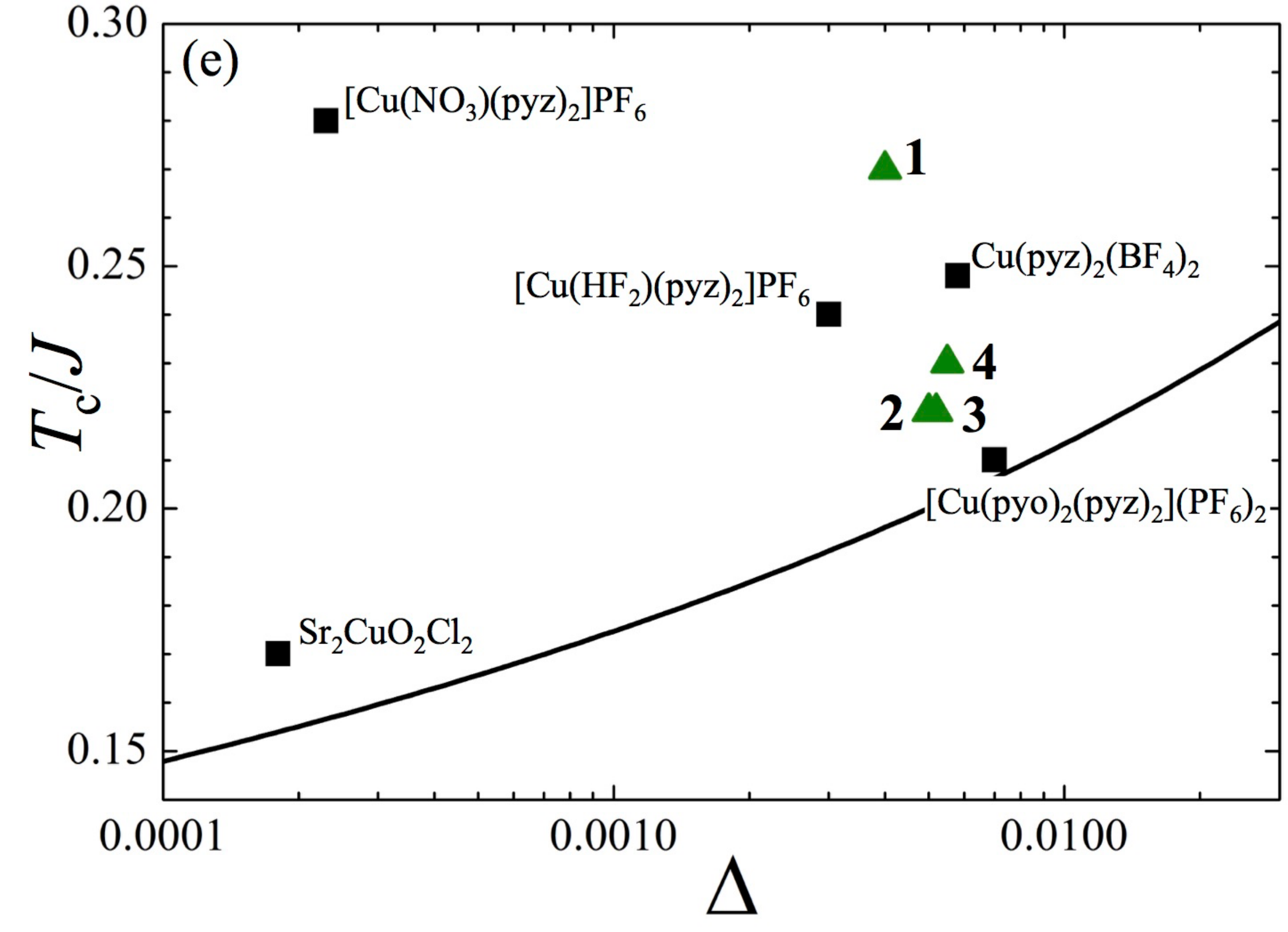}
\sloppypar
\caption{(color online). (a) - (d) Low-field magnetization data (points) for 
polycrystalline samples for {\bf 1}, {\bf 2} and {\bf 3} taken at 
500~mK and a single crystal of compound {\bf 4}. 
For the latter data, taken from Ref.~\onlinecite{fan}, the magnetic field 
was applied both parallel (solid circles) and perpendicular 
(open circles) to the copper-pyrazine planes. 
The anisotropy fields, $H_{\rm A}$, are determined by the peaks 
in ${\rm d}^2M/{\rm d}H^2$ (solid line). 
(e)~Experimentally determined values of $T_{\rm c}/J$ 
and the $XY$-type spin exchange anisotropy, 
$\Delta = H_{\rm A}/H_{\rm c}$, for the compounds 
mentioned in the text (points). 
Also shown is the theoretical relation (line) 
determined via quantum-Monte-Carlo simulations~\cite{cuccoliprb03}.}
\label{kinkfig}
\end{figure}

Low-field magnetization data taken using polycrystalline samples 
of {\bf 1}, {\bf 2} and {\bf 3} are shown in Figure~\ref{kinkfig}(a)-(c). 
These experiments were performed at 500~mK so that the sample 
is in the magnetically ordered state. 
Figure~\ref{kinkfig}(d) shows similar data for {\bf 4} at 
1.8~K taken from~\cite{fan}. 
In this case, a single crystal was used and the magnetic 
field aligned both parallel and perpendicular to the Cu--pyz planes. 
The kink in $M(H)$ arising from the spin-exchange anisotropy is 
clearly observed with the parallel-field orientation. 
This kink occurs when the applied in-plane field is increased above the
anisotropy field, $H_{\rm A}$~[\onlinecite{fan}]. At this point the
moments can begin to rotate out of the easy plane,
leading to an enhancement in the dynamic susceptibility
and hence a kink in the magnetization~\cite{fan}.
The equivalent feature in the data for the polycrystalline 
samples is reduced in clarity by comparison; nevertheless, a change in 
slope is apparent for all compounds. 
The fields at which this kink appears, the anisotropy field 
$\mu_0 H_{\rm A}$, are determined from the position of the large peak in 
${\rm d}^2M/{\rm d}H^2$ and are found to be 
0.08~T ({\bf 1}), 0.11~T ({\bf 2}) and 0.11~T ({\bf 3}). 
$\Delta$ can be found by taking the ratio of $H_{\rm A}$ and 
$H_{\rm c}$~[\onlinecite{fan}] and is found to be similar in magnitude 
($\sim10^{-3}$) for all members of the family. 
The size of the energy scale associated with this anisotropy 
is estimated from the product of $J$ and $\Delta$ and 
these values are tabulated in Table~\ref{magtab}.

If we assume that the polycrystalline compounds also exhibit an 
$XY$-type anisotropy, we can determine the effect this has on the magnetic ordering temperature. Figure~\ref{kinkfig}(e) plots the experimental values of $T_{\rm c}/J$ against $\Delta$ for the four compounds, as well as some other quasi-2D copper-based antiferromagnets. Also shown is the function
\begin{equation}
T_{\rm c}/J = \frac{2.22}{\ln(330/\Delta)},
\end{equation}
which is empirically determined via quantum-Monte-Carlo 
calculations~\cite{cuccoliprb03} and relates the ordering 
temperature to $\Delta$, assuming that the nearest-neighbor, 
anisotropic intralayer exchange is the only term in the 
Hamiltonian (i.e. neglecting $J_2$ and $J_\perp$ in 
Equation~\ref{hamfull}). 
Following Reference~\onlinecite{fan}, we expect that if an experimental 
point lies on this line then the ordering temperature 
can be entirely accounted for by the magnitude of $\Delta$. 
If, however, the point lies significantly above the theoretical curve, 
then the long-range ordering is being assisted by 
another process, such as interlayer coupling. 
As can be seen, the experimentally-derived data
point for {\bf 1}, which has the shortest interlayer 
distance of our four materials, lies some way from 
the theoretical line, while {\bf 4}, with a slightly larger 
interlayer separation, is much closer. 
The points for compounds {\bf 2} and {\bf 3}, with their bulky axial ligands 
and therefore large separation between planes, 
lie close together, not far from the theoretical curve. 
This implies that interlayer coupling 
appears to becomes less important in determining 
the ordering temperature as the copper-pyrazine 
layers are forced apart. 

We can compare our results with those from related materials. 
As has been mentioned, [Cu(HF$_2$)(pyz)$_2$]PF$_6$ has a higher $J$, 
$T_{\rm c}$ and $\Delta$ than [Cu(HF$_2$)(pyz)$_2$]ClO$_4$ ({\bf 1}). 
As a consequence the data point for this material 
lies closer to the theoretical prediction in Figure~\ref{kinkfig}(e) 
than does compound {\bf 1}. 
The measured interlayer separation in the PF$_6$ material is 
6.785~\AA~at 15~K, which is slightly larger than the 
equivalent distance of 6.594~\AA~measured in {\bf 1} at 50~K. 
Also shown in the graph are the points for 
[Cu(pyO)$_2$(pyz)$_2$](PF$_6$)$_2$ and Cu(pyz)$_2$(BF$_4$)$_2$~\cite{kohama10,fan}, 
but as the structures of these compounds are not
known at 100~K or below, a comparison with our materials is difficult. 
The data point for the former material lies right on the theoretical curve. 
Its interlayer separation at room temperature is known to be 
13.189~\AA, compared with 12.471~\AA~at the same 
temperature for {\bf 2}, thus reinforcing the trend. 
However, at low temperatures this separation will certainly not 
be as big as that for {\bf 3} with the large 4-phpyO ligands, 
suggesting that factors beyond just the interlayer separation 
are important for understanding how the different interactions 
determine the magnetic ordering temperature. 

New theoretical models will have to be developed in order 
to achieve an independent estimate of both 
$\Delta$ and $J_\perp$ from the Hamiltonian in Equation~\ref{hamfull}. 
Nevertheless, Figure~\ref{kinkfig}(e) implies that the magnetic 
order observed in compounds {\bf 2} and {\bf 3} is predominantly 
driven by the spin exchange anisotropy. 
In contrast, in order to account for the transition temperature of 
compound {\bf 1}, $\Delta$ would have to be at least an order of 
magnitude larger than is measured. 
Therefore, interlayer couplings, probably including dipolar interactions, 
must play an important role in stabilising the ordered state of this material.

\section{Concluding remarks}
We have engineered a family of molecular materials 
[Cu(HF$_2$)(pyz)$_2$]ClO$_4$ ({\bf 1}), 
[Cu$L_2$(pyz)$_2$](ClO$_4$)$_2$ [$L$ = pyO ({\bf 2}) and 4-phpyO ({\bf 3})] 
that exhibit highly-2D antiferromagnetism. 
By changing the axial ligands we have shown that it is 
possible to vary considerably the interlayer spacings in these 
materials while at the same time retaining the 
copper-pyrazine nearly-square-lattice motif to a very good precision. 
By developing a consistent ESR measurement procedure in 
conjunction with pulsed-field magnetometry, 
we have shown that the ligand substitution has only a 
small effect on the single-ion properties or the magnitude of 
effective nearest-neighbor exchange interactions within the 2D 
layers. Long-range magnetic order is observed in all members 
of the family and the ordering temperatures are found to decrease 
relatively slowly with increasing interlayer separation, 
varying at a rate of  $\sim10$~mK/\AA. 
This slow variation implies that other mechanisms 
in addition to or besides interlayer superexchange 
drive the ordering transition. 
By determining the anisotropy fields in these materials 
we conclude that a combination of spin-exchange 
anisotropy and interlayer coupling 
(superexchange and/or dipolar interactions) gives rise to 
the observed transition temperatures in these materials. 
In particular, the small spin-exchange anisotropy in the 
intralayer exchange interaction becomes increasingly 
more dominant in determining the long-range ordering 
temperature as the layer spacing increases.

Cu(pyz)$_2$(ClO$_4$)$_2$ ({\bf 4}) is a well-known 
molecular antiferromagnet, and has similar copper-pyrazine 
planes to our family of compounds. 
Nevertheless, the intralayer exchange energy $J$, and 
(hence) the magnetic ordering temperature, are 
significantly higher for this material, reminiscent of the 
larger values of $J$ found in some members of the the 
[Cu(HF$_2$)(pyz)$_2$]$X$ family of polymers. 
To fully determine how this difference arises, as well as to 
ascertain in more general terms the means by which exchange 
is mediated through pyrazine and other molecular ligands, 
considerable further work (e.g. density functional theory, 
neutron scattering, spin density mapping) will be required.

Our results imply that characterising the mechanism of the 
in-plane symmetry breaking in real examples of highly 
two-dimensional $S=1/2$ antiferromagnets is necessary 
to achieve a complete understanding of their phase diagram 
and particularly the precise nature of the ordered phases observed 
at very low temperatures (e.g. N\'{e}el, Berezinsky–Kosterlitz–Thouless, 
or 3D XY). It is becoming increasingly clear that similar considerations 
regarding in-plane symmetry are probably vital for explaining the 
high-temperature superconductivity observed in under-doped 
cuprates (see Ref.~\onlinecite{ramshaw2015}
and references therein), whose parent phase is also a quasi-two-dimensional
$S=1/2$ square-lattice antiferromagnet. 
In addition, our data provide guidelines for future attempts to 
create extremely two-dimensional quantum magnets with 
highly suppressed ordering temperatures, thereby engineering a 
magnetic system in close proximity to a quantum critical point 
in the temperature-coupling phase diagram. 
Clearly it is not sufficient to curb the interlayer coupling 
alone; even in $S=1/2$ systems efforts must be taken to 
suppress any intralayer magnetic anisotropy. 
We point out that for the related material [Cu(HF$_2$)(pyz)$_2$]SbF$_6$ it 
was not possible to detect any spin-exchange anisotropy using 
either magnetometry or neutron diffraction down to the lowest
temperature measured ~\cite{brambleby2015},
suggesting that the observed transition to long-range order 
is driven predominantly by interlayer coupling. 
That material, unlike the systems considered here, 
has tetragonal structural symmetry in the ordered phase, 
and it is probable that the magnetic anisotropy we observe 
in our materials is linked to their reduced structural symmetry. 
Whether it is possible to produce a molecule-based material with 
tetragonal symmetry at low temperatures {\sl and} a large 
interlayer spacing is the subject of continuing research.

\section{Acknowledgments}
We thank EPSRC for financial support. 
The work at EWU was supported by the NSF under grant no. DMR-1306158.
A portion of this work was performed at the National High
Magnetic Field Laboratory, which is supported by National
Science Foundation Cooperative Agreement No. DMR-
1157490, the State of Florida, and the U.S. Department of
Energy (DoE) and through the DoE Basic Energy Science
Field Work Proposal ``Science in 100 T''. 
Part of this work was carried out at the STFC ISIS Facility,
Rutheford Appleton Laboratory (UK) and at the Swiss Muon Source,
Paul Scherrer Institut (Switzerland); we are very grateful for
the provision of beamtime. JS thanks the University of
Oxford for provision of a visiting professorship which was 
vital to the completion of this manuscript.

\noindent
$^*$Authors to whom
correspondence should be addressed: p.goddard@warwick.ac.uk;
jmanson@ewu.edu.


\begin{thebibliography}{99}
\bibitem{lines}
M. Lines, Journal of Physics and Chemistry of Solids {\bf 31}, 101 (1970).
\bibitem{chakravarty88}
S. Chakravarty, B. I. Halperin and D. R. Nelson, Phys. Rev. Lett. {\bf 60}, 1057 (1988).
\bibitem{chakravarty89}
S. Chakravarty, B. I. Halperin and D. R. Nelson, Phys. Rev. B {\bf 39}, 2344 (1989).
\bibitem{tyc89}
S. Tyc, B. I. Halperin and S. Chakravarty, Phys. Rev. Lett. {\bf 62}, 835 (1989).
\bibitem{manousakis}
E. Manousakis, Rev. Mod. Phys. {\bf 63}, 1 (1991). 
\bibitem{mermin}
N. D. Mermin and H. Wagner, Phys. Rev. Lett. {\bf 17}, 1133
(1966).
\bibitem{syljuasen02}
O. F. Syljuasen and P. A. Lee, Phys. Rev. Lett. {\bf 88},
207207 (2002). 
\bibitem{cuccoliprb03}
A. Cuccoli, T. Roscilde, V. Tognetti, R. Vaia and P. Verrucchi, 
Phys. Rev. B {\bf 67}, 104414 (2003). 
\bibitem{chandra88}
P. Chandra and B. Doucot, Phys. Rev. B {\bf 38}, 9335 (1988).
\bibitem{sushkov01}
O. P. Sushkov, J. Oitmaa and Z. Weihong, Phys. Rev. B {\bf 63}, 104420 (2001).
\bibitem{njp}
P. A. Goddard, J. Singleton, P. Sengupta, R. D. McDonald, 
T. Lancaster, S. J. Blundell, F. L. Pratt, S. Cox, N. Harrison,
 J. L. Manson, H. I. Southerland and J. A. Schlueter, 
New Journal of Physics {\bf 10}, 083025 (2008).
\bibitem{lancasterprb07}
T. Lancaster, S. J. Blundell, M.L. Brooks, P.J. Baker, F.L. Pratt, 
J.L. Manson, M.M. Conner, F. Xiao, C.P. Landee, 
F.A. Chaves, S. Soriano, M.A. Novak, T.P. Papageorgiou, 
A.D. Bianchi, T. Herrmannsd\"{o}rfer, 
J. Wosnitza and J. A. Schlueter, Phys. Rev. B {\bf 75}, 094421 (2007).
\bibitem{mansonag}
J. L. Manson, K. H. Stone, H. I. Southerland, T. Lancaster, 
A.J. Steele, S.J. Blundell, F.L. Pratt, P. J. Baker, 
R.D. McDonald, P. Sengupta, J. Singleton, P.A. Goddard, 
C. Lee, M.-H. Whangbo, M. M. Warter, 
C. H. Mielke and P. W. Stephens, Journal of the American Chemical Society
{\bf 131}, 4590 (2009).
\bibitem{steele}
A.J. Steele, T. Lancaster, S.J. Blundell, P.J. Baker, 
F.L. Pratt, C. Baines, M.M. Conner, H.I. Southerland, 
J.L. Manson and J.A. Schlueter, Phys. Rev. B {\bf 84}, 064412 (2011).
\bibitem{selmani10}
V. Selmani, C. P. Landee, M. M. Turnbull, J. L. Wikaira and 
F. Xiao, Inorganic Chemistry Communications {\bf 13}, 1399 (2010).
\bibitem{kohama10}
Y. Kohama, M. Jaime, O.E. Ayala-Valenzuela, R.D. McDonald, 
E.D. Mun, J.F. Corbey and J.L. Manson, Phys. Rev. B {\bf 84}, 184402 (2011).
\bibitem{fortune2014}
N.A. Fortune, S.T. Hannahs, C.P. Landee, M.M. Turnbull and F. Xiao, 
J. Phys. Chem. Solids
{\bf 568}, 042004 (2014).
\bibitem{isotope}
P.A. Goddard, J. Singleton, C. Maitland, S.J. Blundell, T. Lancaster, P.J. Baker, 
R.D. McDonald, S. Cox, P. Sengupta, J.L. Manson, K.A. Funk 
and J.A. Schlueter, Phys. Rev. B {\bf 78}, 052408 (2008).
\bibitem{woodward1}
F.M. Woodward, P.J. Gibson, G.B. Jameson, 
C.P. Landee, M.M. Turnbull and R.D. Willett, Inorganic Chemistry 
{\bf 46}, 4256 (2007). 
\bibitem{butcher09}
R.T. Butcher, C.P. Landee, M.M. Turnbull and F. Xiao, 
Inorganica Chemica Acta {\bf 361}, 3654 (2008).
\bibitem{lapidus13}
S.H. Lapidus, J.L. Manson, J. Liu, 
M.J. Smith, P. Goddard, J. Bendix, C.V. Topping, J. Singleton, C. Dunmars, 
J.F. Mitchell and J.A. Schlueter, Chem. Comm. 3558 (2013). 
\bibitem{halder}
G.J. Halder, K.W. Chapman, J.A. Schlueter and J. L. Manson, 
Angewandte Chemie International Edition {\bf 50}, 419 (2011).
\bibitem{saman13}
S. Ghannadzadeh, J. S. Moller, P.A. Goddard, 
T. Lancaster, F. Xiao, S.J. Blundell, A. Maisuradze, 
R. Khasanov, J.L. Manson, S.W. Tozer, D. Graf and 
J.A. Schlueter, Physical Review B {\bf 87}, 241102 (2013).
\bibitem{darriet}
J. Darriet, M. S. Haddad, E. N. Duesler and D. N. Hendrickson, 
Inorganic Chemistry {\bf 18}, 2679 (1979). 
\bibitem{smart} 
SMART: v.5.630, Bruker Molecular Analysis 
Research Tool, Bruker AXS, Madison, WI, 2002.
\bibitem{saintplus}
SAINTPlus: v. 6.45a, Data Reduction and 
Correction Program, Bruker AXS, Madison, WI, 2001.
\bibitem{sadabs1}
SADABS: v.1.05, an empirical absorption 
correction program, Sheldrick, G.M., Bruker AXS Inc., Madison, WI, 2002.
\bibitem{shelxtl}
SHELXTL: v. 6.14, Structure Determination Software 
Suite, Sheldrick, G.M., Bruker AXS Inc., Madison, WI, 2003.
\bibitem{schramaJPCM}
J.M. Schrama, J. Singleton, R.S. Edwards, A. Ardavan, E. Rzepniewski, 
R. Harris, P. Goy, M. Gross, J. Schlueter, M. Kurmoo and P. Day, 
Journal of Physics: Condensed Matter {\bf 13}, 2235 (2001).
\bibitem{noodleman}
L. J. Noodleman, J. Chem. Phys. {\bf 74}, 5737 (1981). 
\bibitem{neese1}
F. Neese, ORCA Version 2.8, revision 2131 (2010). 
\bibitem{neese2}
F. Neese, Coord. Chem. Rev. {\bf 253}, 526 (2009). 
\bibitem{sinnecker}
S. Sinnecker, F. Neese and W. Lubitz, J. Biol. Inorg. Chem. 
{\bf 231}, 10 (2005).
\bibitem{yamaguchi}
K. Yamaguchi, Y. Takahara and T. Fueno, 
Applied Quantum Chemistry, edited by V.H. Smith, 
F. Schafer III and K. Morokuma (D. Reidel: Boston, MA, 1986) p. 155.
\bibitem{soda}
T. Soda, Y. Kitagawa, T. Onishi, Y. Takano, Y. Shigeta, 
H. Nagao, Y. Yoshioka and K. Yamaguchi, Chem. Phys. Lett. {\bf 319}, 223 (2000). 
\bibitem{manson11}
J.L. Manson, S.H. Lapidus, P.W. Stephens, P.K. Peterson, K.E. Carreiro, 
H.I. Southerland, T. Lancaster, S.J. Blundell, A.J. Steele, P.A. Goddard, 
F.L. Pratt, J. Singleton, Y. Kohama, R.D. McDonald, R.E. Del Sesto, 
N.A. Smith, J. Bendix, S.A. Zvyagin, J.H. Kang, C. Lee, 
M.H. Whangbo, V.S. Zapf and A. Plonczak, Inorg. Chem. {\bf 50}, 5990 (2011).
\bibitem{ahlrichs}
A. Schaefer, H. Horn and R. Ahlrichs, J. Chem. Phys. {\bf 97}, 2571 (1992). 
\bibitem{CCDC}
The Cambridge Crystallographic Data Centre (www.ccdc.cam.ac.uk); CCDC reference codes: 683413, 1456465-1456470.
\bibitem{EinM}
Jean-Pierre Launay and Michel Verdaguer, 
{\it Electrons in molecules} (Oxford University Press, Oxford, 2014)
\bibitem{Lancaster2014}
T. Lancaster, P. A. Goddard, S. J. Blundell, F. R. Foronda, 
S. Ghannadzadeh, J. S. M\"{o}ller, P. J. Baker, F. L. Pratt, 
C. Baines, L. Huang, J. Wosnitza, R. D. McDonald, K. A. Modic, 
J. Singleton, C. V. Topping, T. A. W. Beale, F. Xiao, J. A. Schlueter, 
A. M. Barton, R. D. Cabrera, K. E. Carreiro, H. E. Tran and J. L. Manson
Phys. Rev. Lett. {\bf 112}, 207201 (2014)
\bibitem{Abragam}
A. Abragam and B. Bleaney, 
{\it Electron paramagnetic resonance of transition ions},
Oxford University Press (Oxford, 2013).
\bibitem{turovbook}
E.A.Turov, {\it Physical properties of magnetically ordered crystals} (Academic Press, 1965).
\bibitem{Knaflic}
T. Knaflic, M. Klanjsek, A. Sans, P. Adler, M. Jansen, C. Felser and D. Arcon,
Phys. Rev. B {\bf 91}, 174419 (2015).
\bibitem{Dolinsek}
J. Dolinsek, M. Vilfan and S. Zumer (Eds.)
{\it Novel NMR and EPR Techniques},
Lecture Notes in Physics {\bf 684} (Springer, Berlin, 2006).
\bibitem{disclaimer}
Note that previous reports of the $g$-factors for some of these materials have often 
been derived from measurements at only one frequency, 
and/or at temperatures very close to $T_{\rm c}$, resulting in 
slight discrepancies with the values given in the 
present work ({\it c.f.}~\cite{njp,darriet}).
\bibitem{turnbull99}
M.M. Turnbull, A.S. Albrecht, G.B. Jameson, and C.P. Landee, 
Mol. Cryst. and Liq. Cryst. {\bf 335}, 245 (1999). 
\bibitem{godprl}
P. A. Goddard, J. L. Manson, J. Singleton, 
I. Franke, T. Lancaster, A.J. Steele, S.J. Blundell, C. Baines, F.L. Pratt, 
R.D. McDonald, O.E. Ayala-Valenzuela, J.F. Corbey, 
H.I. Southerland, P. Sengupta and J. A. Schlueter, Phys. Rev. Lett. 
{\bf 108}, 077208 (2012). 
\bibitem{yasuda}
C. Yasuda, S. Todo, K. Hukushima, F. Alet, M. Keller, M. Troyer 
and H. Takayama, Phys. Rev. Lett. {\bf 94}, 217201 (2005).
\bibitem{mohri99}
F. Mohri, K. Yoshizawa, T. Yambe, T. Ishida
and T. Nogami, Molecular Engineering {\bf 8}, 357 (1999). 
\bibitem{richardson77}
H.W. Richardson, J. R. Wasson and W. E. Hatfield,
 Inorganic Chemistry {\bf 16}, 484 (1977).
\bibitem{vela13}
S. Vela, J. Jornet-Somoza, M.M. Turnbull, R. Feyerherm, J.J. Novoa 
and M. Deumal, Inorganic Chemistry {\bf 52}, 12923 (2013). 
\bibitem{vela2}
Note the definition of $J$ used in Ref.~\cite{vela13}
 differs by a factor of
2 from the one employed in this paper. 
\bibitem{tsyrulinprl}
N. Tsyrulin, T. Pardini, R. R. P. Singh, F. Xiao, P. Link,
A. Schneidewind, A. Hiess, C. P. Landee, 
M.M. Turnbull and M. Kenzelmann, 
Phys. Rev. Lett. {\bf 102}, 197201 (2009).
\bibitem{viana07}
J. R. Viana and J. R. de Sousa, Phys. Rev. B {\bf 75}, 052403 (2007).
\bibitem{tsyrulinprb}
N. Tsyrulin, F. Xiao, A. Schneidewind, P. Link, H. M. Ronnow, 
J. Gavilano, C.P. Landee, M.M. Turnbull and M. Kenzelmann, 
Phys. Rev. B {\bf 81}, 134409 (2010).
\bibitem{sengupta03}
P. Sengupta, A. W. Sandvik and R. R. P. Singh, Phys. Rev. B {\bf 68}, 094423 (2003).
\bibitem{Blundell99}
S.J. Blundell, Contemporary Physics {\bf 40}, 175 (1999). 
\bibitem{Le1993}
L.P. Le, A. Keren, G.M. Luke, B.J. Sternlieb, W.D. Wu, Y.J. Uemura, J.H. Brewer, 
T.M. Riseman, R.V. Upasani, L.Y. Chiang, W. Kang, 
P.M. Chaikin, T. Csiba and G. Gr\"{u}ner, Phys. Rev. B {\bf 48}, 7284 (1993). 
\bibitem{hastings06}
M. B. Hastings and C. Mudry, Phys. Rev. Lett. {\bf 96},
027215 (2006).
\bibitem{praz06}
A. Praz, C. Mudry, and M. B. Hastings, Phys. Rev. B
{\bf 74}, 184407 (2006).	].
\bibitem{bloch66}
D. Bloch, Journal of Physics and Chemistry of Solids {\bf 27}, 881 (1966).
\bibitem{hutchings68}
M.T. Hutchings, R.J. Birgeneau and W.P. Wolf, Phys. Rev. {\bf 168}, 1026 (1968).
\bibitem{rogers72}
R. N. Rogers, L. Finegol, and B. Morosin, 
Phys. Rev. B {\bf 6}, 1058 (1972).
\bibitem{johnson74}
K. C. Johnson and A. J. Sievers, Phys. Rev. B {\bf 10}, 1027 (1974).
\bibitem{Luttinger1946}
J. M. Luttinger and L. Tisza, Phys. Rev. {\bf 70}, 954 (1946). 
\bibitem{Kraemer2012}
C. Kraemer, N. Nikseresht, J.O. Piatek, N. Tsyrulin, B.D. Piazza, 
K. Kiefer, B. Klemke, T.F. Rosenbaum, G. Aeppli, C. Gannarelli, 
K. Prokes, A. Podlesnyak, T. Stressle, 
L. Keller, O. Zaharko, K. W. Kramer and
H.M. Ronnow, Science {\bf 336}, 1416 (2012). 
\bibitem{kotler14}
S. Kotler, N. Akerman, N. Navon, Y. Glickmann and
R. Ozeri, Nature {\bf 510}, 376 (2014).
\bibitem{kurmooferro}
Kurmoo~\cite{kurmoo99} has shown that dipole 
interactions are responsible for the ordering in the family of 
quasi-2D materials based on cobalt hydroxide layers, 
which have interlayer separations as large as 22.8~\AA. 
In those materials the intralayer exchange interactions are 
ferrimagnetic in nature and so the onset of short-range order causes 
a significant moment to develop within the layers, 
thus enhancing the effectiveness of the dipolar coupling.
\bibitem{kurmoo99} 
M. Kurmoo, Chem. Mater. {\bf 11}, 3370 (1999). 
\bibitem{fan}
F. Xiao, F.M. Woodward, C.P. Landee, M.M. Turnbull,
C. Mielke, N. Harrison, T. Lancaster, S.J. Blundell, P J. Baker, 
P. Babkevich and F.L. Pratt, Phys. Rev. B {\bf 79}, 134412 (2009).
\bibitem{dingprl92}
H.-Q. Ding, Phys. Rev. Lett. {\bf 68}, 1927 (1992). 
\bibitem{LaterESR}
A later ESR study provided evidence for a further, 
smaller anisotropy within the $xy$-plane in compound {\bf 4}~[\onlinecite{povarov}].
\bibitem{povarov}
K. Yu. Povarov, A.I. Smirnov and C. P. Landee, Phys. Rev. B 
\textbf{87}, 214402 (2013).
\bibitem{cizmar10}
E. Cizmar, S.A. Zvyagin, R. Beyer, M. Uhlarz, M. Ozerov, Y. Skourski, 
J.L. Manson, J.A. Schlueter and J. Wosnitza, Phys. Rev. B {\bf 81}, 064422 (2010).
\bibitem{ramshaw2015}
B.J. Ramshaw, S.E. Sebastian, R.D. McDonald, 
James Day, B. S. Tan, Z. Zhu, J. B. Betts, 
Ruixing Liang, D.A. Bonn, W.N. Hardy and N. Harrison,
Science {\bf 348}, 317 (2015).
\bibitem{brambleby2015}
J. Brambleby, P.A. Goddard, R.D. Johnson, J. Liu, 
D. Kaminski, A. Ardavan, A.J. Steele, S.J. Blundell, 
T. Lancaster, P. Manuel, P.J. Baker, J. Singleton, 
S.G. Schwalbe, P.M. Spurgeon, H.E. Tran, P.K. Peterson,
 J.F. Corbey and J.L. Manson, Phys. Rev. B {\bf 92}, 134406 (2015).
\end{thebibliography}
\end{document}